\documentclass[preprint,showpacs,preprintnumbers,amsmath,amssymb]{revtex4}
\usepackage{amsmath}
\usepackage{amsthm}
\usepackage{amssymb}
\usepackage[german,english]{babel}
\usepackage[dvips,unicode]{hyperref}

\newtheorem{prop}{Proposition}
\newtheorem*{rem}{Remark}

\begin{document}
\title{Multi-Component Gas-Dynamics and Turbulence}

\author{S.~A.~Serov}
 \email{serov@vniief.ru}
 \affiliation{Russian Federal Nuclear Centre --
   All-Russian Scientific Research Institute of Experimental Physics,
   Institute of Theoretical and Mathematical Physics, Sarov, Russia}
\author{S.~S.~Serova}
 \affiliation{Physics Faculty, St. Petersburg State University, 
   St. Petersburg, Russia}

\date{\today}

\begin{abstract}
In the article, correct method for the kinetic Boltzmann equation
asymptotic solution is formulated,
the Hilbert method and the Enskog error are considered.
The equations system of multi-component nonequilibrium gas-dynamics 
is derived, that corresponds to the first order in the approximate
method for solution of the kinetic Boltzmann equation within the Struminskii approach.
It is shown, that velocity distribution functions, received by the
proposed method and by the
Enskog method within the Enskog approach to solving of kinetic
Boltzmann equation for gas mixture, 
are equivalent to first infinitesimal order terms 
(inclusive; accordingly, systems of gas-dynamic equations of the second order coincide),
but, generally speaking, differ in the next order.
An interpretation of turbulent gas flow is proposed, as stratified 
to components gas flow, which is described by the derived equations
system of multi-component nonequilibrium gas-dynamics.
\end{abstract}


\maketitle

\section{Introduction}
\label{sec:introduction}

In 1912 Hilbert in \cite{hilbert12}, Chapter~XXII, as an example of
integral equation, considered the kinetic Boltzmann equation for
one-component gas and proposed a "recipe" for its approximate
(asymptotic) solution.
Enskog concretized the Hilbert "recipe".
But meantime Enskog had made a not obvious logical mistake
(see below section \ref{sec:remarks}) and, as the result, had formulated untrue
method for (asymptotic) solution of the kinetic Boltzmann equation, having
proposed to use "null" conditions, conditions
(\ref{r_condition1})-(\ref{r_condition3n}) below with zero right-hand
side, for determination of five arbitrary functional parameters of the
first and following approximations of the velocity distribution function.
As result of paralogism of the successive approximations method, 
partial time derivatives vanish in
the necessary and sufficient conditions of solutions existence of integral
equations of higher orders, see equations (\ref{system_r}) below, and with them
terms of gas-dynamic equations, corresponding viscosity, heat 
conductivity \ldots , vanish.
Enskog "improved" the situation by the introduction 
(see, for example, \cite{chap60}, Chapter~7, \S ~1, 
Section~5) of the unreasonable expansion of partial time derivative.

Below, on the example of proposed by Struminskii in \cite{strum74}
method for approximate solution of the kinetic Boltzmann equation for
multi-component gas, it is shown, how it is necessary to change
the Enskog method.
The Struminskii approach differs from the Enskog approach to
approximate solving of the kinetic Boltzmann equation for gas mixture (see, for
example, \cite{chap60}, Chapter~8, \S ~2) in the expansion of the Boltzmann
equation -- see (\ref{epsstrum}) below.
It should be noted, that approaches to 
the approximate solving of the kinetic Boltzmann equation, that are close to 
that suggested by Struminskii, were considered previously in the kinetic 
theory of plasmas, see, e.g., \cite{gurov66}, \S ~7.5. 
Struminskii's paper \cite{strum74} and, 
hence, his following papers, referring to \cite{strum74} (for example,  
\cite{strum82}), involves also errors in calculation of collision integrals. 
This led Struminskii to improper conclusions of general nature.
Proposed change of the Struminskii method does not clear up the 
principal demerit of the Struminskii's method, i.e. the lack of the
explicit physical small parameter, in terms of which the expansion of the successive 
approximations method is performed (but it is not very significant for the 
content of this article, as we shall see below in the section
\ref{sec:order1}).
 
Further in this article calculations of collision integrals of the general form 
and for a specific model of the rigid sphere potential for 
the case, where separate components have, generally speaking, different mean 
velocities and temperatures, are presented. 
The equations system of multi-component nonequilibrium gas-dynamics 
is derived, that corresponds to the first order in the approximate
method for solution of the kinetic Boltzmann equation (the approximation
order is defined below) within the Struminskii approach.
It is shown, that velocity distribution functions, received by the
described method and by the
Enskog method within the Enskog approach to solving of kinetic
Boltzmann equation for gas mixture, 
are equivalent to first infinitesimal order terms 
(inclusive; accordingly, systems of gas-dynamic equations of the second order coincide),
but, generally speaking, differ in the next order.
This difference is the possible reason, that going to higher order 
in the Enskog method does not 
lead to any real improvement in the results. 

At the end of this article an interpretation of turbulent flows is proposed within 
the multi-component gas-dynamics. 

The below notation is close to the one in \cite{chap60}; this allows an 
easy comparison of the described below theory to the Enskog-Chapman theory 
and replacement of the treatment of details, common to the theories, by 
references to the appropriate points in \cite{chap60}. 
Also it may be useful a familiarity with basics of the
asymptotic expansions theory, accurately stated, for
example, in \cite{bourbaki2004}, Chapter~V.

\section{Modified Struminskii's method for kinetic Boltzmann equation solution}
\label{sec:method}

The basic idea of the Struminskii's approach is as follows. 
In Boltzmann equation for mixture of rarefied monatomic gases 
[see \cite{chap60}, Chapter~8, (1.1); the derivation 
of the Boltzmann equation and its domain of applicability are discussed, for example, 
in \cite{chap60}, Chapter~3 and 18, \cite{hirsch61}, Chapter~7, \S ~1 
and the Bogolyubov paper \cite{bog46}, 
which is included in \cite{chap60} as an addition]:
\begin{eqnarray}
\frac{\partial f_{i}}{\partial t}&+&
{\mathbf{c}}_{i}\cdot \frac{\partial f_{i}}{\partial \mathbf{r}}
+ \frac{{\mathbf{X}}_{i}}{m_{i}}\cdot \frac{\partial f_{i}}
{\partial {\mathbf{c}}_{i}} \nonumber\\*
&&=
\sum\limits_{j} \iiint
\left( f_{i}^{\prime }f_{j}^{\prime }-f_{i}f_{j} \right)
g_{ij} b\,db \,d\epsilon \,d{\mathbf{c}}_{j} \nonumber\\*
&&=
\sum\limits_{j} \iint
\left( f_{i}^{\prime }f_{j}^{\prime }-f_{i}f_{j} \right)
k_{ij} \,d{\mathbf{k}} \,d{\mathbf{c}}_{j} ,
\label{boltzmannend}
\end{eqnarray}
the formal parameter $\theta $ 
(used simply as an "indicator of smallness", $\theta =1$)
is included in somewhat another manner, than 
in the Enskog approach:
\begin{eqnarray}
{{\mathcal D}_i}{f_i} =
- \frac{1}{\theta }\ J_{i}\left( f_{i},f\right)
- \sum\limits_{j\neq i}J_{ij}\left( f_{i},f_{j}\right) .
\label{epsstrum}
\end{eqnarray}
In the Enskog approach multiplier $1\left/ {\theta}\right.$ refers to the whole 
right-hand side:
\begin{eqnarray}
{{\mathcal D}_i}{f_i} =
- \frac{1}{\theta }\sum\limits_{j}J_{ij}\left( f_{i},f_{j}\right) 
\label{epsEnskog}
\end{eqnarray}
-- see \cite{chap60}, Chapter~7, \S ~1, Section~5. 

In (\ref {boltzmannend})-(\ref {epsstrum}) 
subscripts $_{i, j}$ number components of the mixture;
${\mathbf{X}}_{i}$ -- the external force, acting on molecule of $i$-th grade; 
$m_{i}$ -- the mass of molecule of $i$-th grade;
$g_{ij}$ -- the module of relative velocity 
${\mathbf{g}}_{ij}={\mathbf{c}}_{i}-{\mathbf{c}}_{j}$ 
of colliding particles;
$b$ -- aiming distance, $\epsilon$ -- azimuthal angle,  
${\mathbf{k}}$ is unit vector directed to the center of 
mass of colliding particles from the point of their closest approach to each 
other, see \cite{chap60}, Chapter~3, figure~3;
the scalar function $k_{ij} \left( {{\mathbf{g}}_{ij} ,{\mathbf{k}}} \right)$ 
is defined by the equality
\begin{eqnarray}
g_{ij} b\,d b \,d \epsilon 
\ {\overset{\mathrm{def}}{=}}\ 
k_{ij} \,d {\mathbf{k}} ;
\label{def_k_ij}
\end{eqnarray}
in (\ref {epsstrum})
\begin{eqnarray}
J_{i}\left( f_{i},f\right) &=&
\iint
\left( f_{i}f-f_{i}^{\prime }f^{\prime } \right)
k_{i} \,d {\mathbf{k}} \,d {\mathbf{c}} ,
\label{J_i}
\\
J_{ij}\left( f_{i},f_{j}\right) &=& 
\iint
\left( f_{i}f_{j}-f_{i}^{\prime }f_{j}^{\prime } \right)
k_{ij} \,d {\mathbf{k}} \,d {\mathbf{c}}_{j} ,
\label{J_ij}
\end{eqnarray}
to differentiate between velocities of colliding molecules of the 
same grade in (\ref{J_i}), we will denote one velocity by ${\mathbf{c}}_i $ and the 
other by ${\mathbf{c}}$ (without any subscript) and omit the subscript in the 
relevant \textit{velocity distribution function} $f$ -- 
cf. with \cite{chap60}, Chapter~3, \S ~5, Section~1;
the other notation is essentially the same as in 
\cite {chap60}, Chapter~3, \S ~5 and Chapter~8, \S ~2. 

Formally introduce parameter $\theta $ to the series of successive 
approximations for the velocity distribution function $f_{i} $: 
\begin{eqnarray}
f_{i}=f_{i}^{\left( 0\right)}+\theta f_{i}^{\left( 1\right)}+
\theta ^{2}f_{i}^{\left( 2\right)}+\dotsb .
\label{f_series}
\end{eqnarray}
Write the differential part of equation (\ref{boltzmannend}) as: 
\begin{eqnarray}
{\mathcal D}_i f_i  &=& 
\left( { \frac{\partial }
{\partial t} + {\mathbf{c}}_{i}\cdot \frac{\partial }{\partial \mathbf{r}} + 
\frac{{\mathbf{X}}_{i}}{m_{i}}\cdot \frac{\partial }
{\partial {\mathbf{c}}_{i}}} \right)\left( {f_i^{\left( 0 \right)}  + 
\theta f_i^{\left( 1 \right)}  + \dotsb } \right) \nonumber\\*
&=& 
{\mathcal D}_i^{\left( 1 \right)}  + \theta {\mathcal D}_i^{\left( 2 \right)}  
+ \theta ^2 {\mathcal D}_i^{\left( 3 \right)}  + \dotsb , 
\label{Dseries}
\\
{\mathcal D}_i^{\left( r \right)}
&=&\frac{\partial f_{i}^{\left( r-1\right)}}{\partial t}+
{\mathbf{c}}_{i}\cdot \frac{\partial 
f_{i}^{\left( r-1\right)}}{\partial \mathbf{r}} +
\frac{{\mathbf{X}}_{i}}{m_{i}}\cdot \frac{\partial 
f_{i}^{\left( r-1\right)}}
{\partial {\mathbf{c}}_{i}} ,
\label{D_r}
\end{eqnarray}
-- cf. with \cite{chap60}, Chapter~7, \S ~1, Sections~4,~5 and \cite{strum74}.
In (\ref{Dseries})-(\ref{D_r}) the partial 
time derivative expansion
\begin{eqnarray}
\frac{\partial }
{{\partial t}} =
\sum\limits_{r=0}^\infty {\theta ^r } \frac{{\partial _r }}
{{\partial t}} ,
\label{t_series_Enskog}
\end{eqnarray}
used by Enskog and then by Struminskii, is not. 
As a result, the below-discussed method for the Boltzmann equation 
solution differs in essence from the Enskog's and Struminskii's 
methods. 

Having substituted (\ref{f_series}) and (\ref{Dseries}) into (\ref{epsstrum}) 
and equated coefficients in like 
powers of $\theta $, we arrive at the system of equations of the successive 
approximations method for finding of functions 
$f_{i}^{\left( {r} \right)} $. 
This system of equations can be written as 
[cf. with \cite{chap60}, Chapter 8, (2.5)-(2.8)]: 
\begin{eqnarray}
{\mathcal D}_i^{\left( r \right)} +
\sum\limits_{j \ne i} J_{ij}^{\left( r \right)} +
{\hat J}_i^{\left( r \right)} + {\check J}_i^{\left( r \right)} = 0 ,
\label{sys_eq_series}
\end{eqnarray}
where
\begin{eqnarray}
{\mathcal D}_i^{\left( 0 \right)} &=& 0 ,
\label{sys_eq_series_D0}
\\
J_{ij}^{\left( 0 \right)} &=& 0 ,
\label{sys_eq_series_ij0}
\\
{\hat J}_i^{\left( 0 \right)} &=& 0 ,
\label{sys_eq_series_ih0}
\\
{\check J}_i^{\left( 0 \right)} &=&
 J_i \left( {f_i^{\left( 0 \right)} f^{\left( {0} \right)} } \right) +
 J_i \left( {f_i^{\left( {0} \right)} f^{\left( 0 \right)} } \right) ,
\label{sys_eq_series_ic0}
\\
{\hat J}_i^{\left( 1 \right)} &=& 0 ,
\label{sys_eq_series_ih1}
\end{eqnarray}
\begin{eqnarray}
\!\!\!\!\!\!\!J_{ij}^{\left( r \right)} &=&
 J_{ij} \left( {f_i^{\left( 0 \right)} f_j^{\left( {r - 1} \right)} } \right) + \dots  +
 J_{ij} \left( {f_i^{\left( {r - 1} \right)} f_j^{\left( 0 \right)} } \right) ,
\label{sys_eq_series_ijr}
\\
\!\!\!\!\!\!\!{\hat J}_i^{\left( r+1 \right)} &=&
 J_i \left( {f_i^{\left( 1 \right)} f^{\left( {r} \right)} } \right) + \dots  +
 J_i \left( {f_i^{\left( {r} \right)} f^{\left( 1 \right)} } \right) ,
\label{sys_eq_series_ihr}
\\
\!\!\!\!\!\!\!{\check J}_i^{\left( r \right)} &=&
 J_i \left( {f_i^{\left( 0 \right)} f^{\left( {r} \right)} } \right) +
 J_i \left( {f_i^{\left( {r} \right)} f^{\left( 0 \right)} } \right) ,
\label{sys_eq_series_icr}
\end{eqnarray}
$r=1,2\ldots $ .

The assumption that ${\mathcal D}_i^{\left( r \right)}$ does not contain 
$f_{i}^{\left({r} \right)} $, reduces the problem of the integro-differential 
equation solution to a simpler problem of the integral equation solution. 

The formal parameter $\theta $ is not introduced to the Boltzmann equation 
in the Enskog approach (\cite{chap60}, Chapter~7, \S ~1, Section~5), 
but the $\theta $ is introduced to the series of 
successive approximations for the velocity distribution functions in another 
manner than in (\ref{f_series}): 
\begin{eqnarray}
f_{i}=\frac{1}{\theta }\,f_{i}^{\left( 0\right)}+
f_{i}^{\left( 1\right)}+
\theta f_{i}^{\left( 2\right)}+ \dotsb .
\label{f_seriesEnscog}
\end{eqnarray}
The result is the same, if expansion (\ref{f_series}) is used for the 
velocity 
distribution function, but multiplier $1 \left/ {\theta} \right.$ referring to 
the whole right-hand side is introduced to equation (\ref{epsEnskog}). 
The successive approximations 
$f_{i}^{\left( 0 \right )}, f_{i}^{\left( 1 \right)}, f_{i}^{\left( 2 \right )} 
\ldots $ calculated within the Enskog approach prove 
ordered in (inverse) mixture \textit{molecule number density} $n$: 
$f_{i}^{\left( 0 \right )}$ is proportional to $n$, 
$f_{i}^{\left( 1 \right )}$ is 
independent of $n$, etc., -- see \cite{chap60}, Chapter~7, \S ~1, Section~5 and 
\cite{chap60}, Chapter~7, \S ~2. 
Ipso facto, a physical substantiation of the using of the successive 
approximations method for finding of a solution of the kinetic 
Boltzmann equation appears. 
For Struminskii's approach it is inconveniently to define explicitly 
a small physical parameter, in terms of which the expansion proceeds. 

Below, when discussing the approximation order, we will consider the 
approximation order equal to the value of the superscript $r$ in 
(\ref{sys_eq_series}). 
In the zeroth approximation we obtain the following integral equation for 
determination of function $f_{i}^{\left( 0 \right)}$: 
\begin{eqnarray}
J_{i}\left( f_{i}^{\left( 0\right)},f^{\left( 0\right)}\right) = 0 .
\label{system0_eq}
\end{eqnarray}
Having multiplied equation (\ref {system0_eq}) by 
$ \ln f_{i}$, integrated  over all ${\mathbf{c}}_{i}$, and 
transformed the integrals with taking into account 
\begin{eqnarray}
\iiint \phi _{i}f_{i}^{\prime }f_{j}^{\prime }
k_{ij} \,d {\mathbf{k}} \,d \mathbf{c}_{i}\,d {\mathbf{c}}_{j}
=\iiint \phi _{i}^{\prime }f_{i}f_{j}
k_{ij} \,d {\mathbf{k}} \,d \mathbf{c}_{i}\,d {\mathbf{c}}_{j} ,
\label{equality}
\end{eqnarray}
with equality (\ref {equality}) being independent of the form of functions 
$\phi ,f$ (cf. with \cite{chap60}, Chapter~3, \S ~3, Section~3), similarly to 
\cite {chap60}, Chapter~4, \S ~3, we find: 
\begin{eqnarray}
\frac{1}{4}
\iiint
\ln \left( 
\frac
{f_i^{\left( 0\right)} f^{\left( 0\right)}}
{f_{i}^{\left( 0\right) \prime }f^{\left( 0\right) \prime }} 
\right)
\left( f_{i}^{\left( 0\right) \prime }f^{\left( 0\right) \prime }-
f_{i}^{\left( 0\right)}f^{\left( 0\right)}\right)
k_{i} \,d {\mathbf{k}} \,d \mathbf{c}\,d {\mathbf{c}}_{i}=0 .
\label{H-equality}
\end{eqnarray}
The integrand in (\ref {H-equality}) cannot be greater than zero, therefore the 
integral in (\ref {H-equality}) can be equal to zero only when the integrand 
(it is supposed, that all considered functions are continuous in each point of 
the range of their definition) vanishes for all
values of the variables, i.e. 
\begin{eqnarray}
f_{i}^{\left( 0\right) \prime }f^{\left( 0\right) \prime }
= f_{i}^{\left( 0\right)}f^{\left( 0\right)}
\label{res_H-equality}
\end{eqnarray}
or
\begin{eqnarray}
\ln f_{i}^{\left( 0\right)}+\ln f^{\left( 0\right)} -
\ln f_{i}^{\left( 0\right) \prime }-\ln f^{\left( 0\right) \prime } = 0 .
\label{res_ln_H-equality}
\end{eqnarray}
Hence, $\ln f_{i}^{\left( 0\right)}$ should be expressed linearly  in terms 
of \textit{additive collision invariants}
$\psi _{i}^{\left( 1\right)}= m_i$,
${\boldsymbol{\psi}}_{i}^{\left( 2\right)}= m_{i}{\mathbf{c}}_{i}$,
$\psi _{i}^{\left( 3\right)}= \frac{1}{2}\, m_{i}{\mathbf{c}}_{i}^{2}$.

For the collision of the $i$-th molecule with the $j$-th molecule the 
conservation of invariant $\psi_{i}^{\left( l\right)}$ is expressed by 
equality: 
\begin{eqnarray}
\psi _{i}^{\left( l\right) \prime }+ \psi _{j}^{\left( l\right) \prime }-
\psi _{i}^{\left( l\right)}-\psi _{j}^{\left( l\right)} = 0 
\qquad {\left( l=1,2,3\right)} .
\label{add_inv}
\end{eqnarray}
As molecule velocities $\mathbf{c}_{i}^{\prime }$,~$\mathbf{c}_{j}^{\prime }$
upon the collision are determined completely in terms of the molecule velocities 
$\mathbf{c}_{i}$,~$\mathbf{c}_{j}$ before the collision by two parameters specifying the 
collision, for example, by aiming distance $b$ and azimuthal angle 
$\epsilon$ (see above), from four scalar equations (\ref{add_inv}), 
corresponding to 
conservation of energy and three components of momentum, like in 
\cite{chap60}, Chapter~3, \S ~2 
we conclude, that no additive collision invariants linearly 
independent of $\psi_{i}^{\left( l\right)}$ exist. However, this 
conclusion seems not quite stringent logically.

Thus,
\begin{eqnarray}
\ln f_{i}^{\left( 0\right)}=\alpha _{i}^{\left( 1,0\right)}+
\boldsymbol{\alpha }_{i}^{\left( 2,0\right)}\cdot m_{i}\mathbf{c}_{i}-
\alpha _{i}^{\left( 3,0\right)}\frac{1}{2}\,m_{i}c_{i}^{2} ,
\label{ln_f_1}
\end{eqnarray}
where $\alpha _{i}^{\left( 1,0 \right)}$ and $\alpha _{i}^{\left( 3,0\right)}$ 
are some scalar functions of $\mathbf{r}$ and $t$, independent of $\mathbf{c}_{i}$, and 
$\boldsymbol{\alpha} _{i}^{\left( 2,0\right)}$ is a vector function of $\mathbf{r}$ and $t$. 
Or 
\begin{eqnarray}
\ln f_{i}^{\left( 0\right)}=\ln \alpha _{i}^{\left( 0,0\right)}-
\alpha _{i}^{\left( 3,0\right)} \frac{1}{2}\,m_{i}
\left( \mathbf{c}_{i}-
\frac
{\boldsymbol{\alpha }_{i}^{\left( 2,0\right)}}
{\alpha _{i}^{\left( 3,0\right)}}
\right) ^{2} ,
\label{ln_f_0}
\end{eqnarray}
where $\alpha _{i}^{\left( 0,0\right)}$ is a new scalar function of $\mathbf{r}$ 
and $t$. 
I.e. the general solution of equation (\ref{system0_eq}) can be written in the 
form of Maxwell function: 
\begin{eqnarray}
f_{i}^{\left( 0\right)}=
\beta _{i}^{\left( 1\right)}
\left( \frac{m_{i}}{2\pi k\beta _{i}^{\left( 3\right)}}\right) ^{\frac{3}{2}}
e^{-{\left. m_{i}\left( {\mathbf{c}}_{i}-
\boldsymbol{\beta }_{i}^{\left( 2\right)}\right) ^{2} \right/ 
{2k\beta _{i}^{\left( 3\right)}}}} ,
\label{maxwell_f0}
\end{eqnarray}
where 
\begin{eqnarray}
\beta _{i}^{\left( 1\right)}&=&
\alpha _i^{\left( {0,0} \right)} \left( {\frac{{2\pi }}
{{m_i \alpha _i^{\left( {3,0} \right)} }}} \right)^{\frac{3}{2}}, 
\label{beta1}
\\
\boldsymbol{\beta }_{i}^{\left( 2\right)} &=&
\frac{{\boldsymbol{\alpha }_{i}^{\left( 2,0\right)}}}
{{\alpha _i^{\left( {3,0} \right)} }},
\label{beta2}
\\
\beta _i^{\left( 3 \right)}  &=& 
\frac{1}{{k\alpha _i^{\left( {3,0} \right)} }}.
\label{beta3}
\end{eqnarray}

By definition particle number density, mean velocity and 
temperature of the $i$-th component are introduced: 
\begin{eqnarray}
n_{i} &{\overset{\mathrm{def}}{=}}& 
\int f_{i}\,d {\mathbf{c}}_{i} ,
\label{condition1}
\\
n_{i} m_{i} {\mathbf{u}}_{i} &{\overset{\mathrm{def}}{=}}& 
\int m_{i} {\mathbf{c}}_{i} f_{i}\,d {\mathbf{c}}_{i} ,
\label{condition2}
\\
\frac{3}{2}\, n_{i} kT_{i} &{\overset{\mathrm{def}}{=}}& 
\int \frac{1}{2}\, m_{i} 
\left( {\mathbf{c}}_{i}-{\mathbf{u}}_{i}\right) ^{2} f_{i}
\,d {\mathbf{c}}_{i} .
\label{condition3}
\end{eqnarray}
From (\ref{condition1})-(\ref{condition3}) we obtain equality:
\begin{eqnarray}
\frac{3}{2}\,n_{i} kT_{i} + 
\frac{1}{2}\, n_{i} m_{i} {u}_{i} ^{2} \, =
\int \frac{1}{2}\, m_{i} {c}_{i}^{2} f_{i} \,d {\mathbf{c}}_{i},
\label{condition3n}
\end{eqnarray}
which is convenient for using further instead of definition
(\ref{condition3}).

Together with asymptotic expansion (\ref {f_series}), 
according to definitions (\ref {condition1}), (\ref {condition2}), 
(\ref {condition3n}), \textit {it is necessary} to define 
asymptotic expansion for particle number density $n_{i}$ of the $i$-th component
\begin{eqnarray}
n_{i}=n_{i}^{\left( 0\right)}+\theta n_{i}^{\left( 1\right)}+
\theta ^{2}n_{i}^{\left( 2\right)}+\dotsb ,
\label{n_series}
\end{eqnarray}
mean velocity ${\mathbf{u}}_{i}$ of the $i$-th component
\begin{eqnarray}
{\mathbf{u}}_{i}={\mathbf{u}}_{i}^{\left( 0\right)}+
\theta {\mathbf{u}}_{i}^{\left( 1\right)}+
\theta ^{2}{\mathbf{u}}_{i}^{\left( 2\right)}+\dotsb 
\label{u_series}
\end{eqnarray}
and temperature $T_{i}$ of the $i$-th component
\begin{eqnarray}
T_{i}=T_{i}^{\left( 0\right)}+\theta T_{i}^{\left( 1\right)}+
\theta ^{2}T_{i}^{\left( 2\right)}+\dotsb .
\label{T_series}
\end{eqnarray}

Having substituted (\ref{f_series}) and (\ref{n_series})-(\ref{T_series}) into 
(\ref{condition1}), (\ref{condition2}), (\ref{condition3n}) and 
equated zeroth infinitesimal order terms,
arbitrary functions 
$\beta _{i}^{\left( 1\right)}\left( {\mathbf{r}},t\right)$,
$\boldsymbol{\beta }_{i}^{\left( 2\right)}\left(
  {\mathbf{r}},t\right)$ and 
$\beta _{i}^{\left( 3\right)}\left( {\mathbf{r}},t\right)$,
that appear in (\ref{maxwell_f0}), can be juxtaposed with the zeroth 
approximations to 
local values of particle number density, mean velocity, and 
temperature of the $i$-th component:
\begin{subequations}
 \label{param_f}
 \begin{eqnarray}
  \beta _{i}^{\left( 1\right)}\left( {\mathbf{r}},t\right) 
&=&
  n_{i}^{\left( 0\right)}\left( {\mathbf{r}},t\right),
\label{param_fa} \\
  \boldsymbol{\beta }_{i}^{\left( 2\right)}\left(
    {\mathbf{r}},t\right) 
&=& {\mathbf{u}}_{i}^{\left( 0\right)}
  \left( {\mathbf{r}},t\right) ,
\label{param_fb} \\
  \beta _{i}^{\left( 3\right)}\left( {\mathbf{r}},t\right) 
&=& T_{i}^{\left( 0\right)}\left( {\mathbf{r}},t\right).
\label{param_fc}
 \end{eqnarray}
\end{subequations}

Functions (\ref{param_f}) are found from 
the first-order gas-dynamic equations system (below we will see, why 
from this equations system): 
\begin{eqnarray}
\int \psi _{i}^{\left( l\right)} 
\left( {\mathcal D}_i^{\left( 1 \right)} +
\sum\limits_{j \ne i} J_{ij}^{\left( 1 \right)} \right)
d {\mathbf{c}}_{i}=0 
\qquad {\left( l=1,2,3\right)} .
\label{gas_dyn_system1}
\end{eqnarray}

Similarly, functions $f_{i}^{\left( r\right)}$ ($r=1,2\ldots $), found at the $r$-th step of the 
successive approximations method, prove, see below, parametrically 
dependent on 5 arbitrary scalar functions of $\mathbf{r}$,~$t$. 
Having equated terms of the same infinitesimal order in 
(\ref{condition1}), (\ref{condition2}), (\ref{condition3n}):
\begin{eqnarray}
\int f_{i}^{\left( r\right)}\,d {\mathbf{c}}_{i}&=&n_{i}^{\left( r\right)} ,
\label{r_condition1}
\\
\int
m_{i} {\mathbf{c}}_{i}f_{i}^{\left( r\right)}\,d {\mathbf{c}}_{i}
&=&
m_{i}\, {\left( n_i {\mathbf{u}}_{i}\right)}^{\left( {r} \right)}
=
m_{i} \sum\limits_{s = 0}^r 
{n_i^{\left( r - s \right)} } {\mathbf{u}}_{i}^{\left( {s} \right)} ,
\label{r_condition2}
\\
\int
\frac{1}{2}\, m_{i} {c}_{i}^{2}
f_{i}^{\left( r\right)}\,d {\mathbf{c}}_{i}
&=&
\frac{3}{2}\, k\, {\left( n_{i} T_{i} \right)}^{\left( {r} \right)}
+ \,\frac{1}{2}\, m_{i} {\left( n_{i} u_{i}^{2} \right)}^{\left( {r} \right)}
\nonumber\\
&=&
\frac{3}{2}\, k \sum\limits_{s = 0}^r 
{n_i^{\left( r - s \right)} } {T_{i}^{\left( {s} \right)}}
+ \,\frac{1}{2}\, m_{i} 
\sum\limits_{s = 0}^{r} \sum\limits_{q = 0}^{s} 
n_{i}^{\left( {r-s}\right)}
{\mathbf{u}}_{i}^{\left( {s-q} \right)}\cdot {\mathbf{u}}_{i}^{\left( {q} \right)},
\label{r_condition3n}
\end{eqnarray}
we obtain 5 conditions for each subscript $i$, that can be used to 
express 5 arbitrary functional parameters in $f_{i}^{\left( r\right)}$ as
functions of 
\begin{subequations}
 \label{param_f_r}
 \begin{eqnarray}
  n_{i}^{\left( r\right)} &=&
  n_{i}^{\left( r\right)}\left( {\mathbf{r}},t\right),
  \label{param_f_ra} \\
  {\mathbf{u}}_{i}^{\left( r\right)} &=& {\mathbf{u}}_{i}^{\left( r\right)}
  \left( {\mathbf{r}},t\right) ,
  \label{param_f_rb} \\
  T_{i}^{\left( r\right)} &=& T_{i}^{\left( r\right)}\left( {\mathbf{r}},t\right),
  \label{param_f_rc}
 \end{eqnarray}
\end{subequations}

Functions (\ref{param_f_r}) can be found from the ${\left( r+1\right)}$-th 
order equations system: 
\begin{eqnarray}
\int \psi _{i}^{\left( l\right)}
\left( {\mathcal D}_i^{\left( r+1 \right)} +
\sum\limits_{j \ne i} J_{ij}^{\left( r+1 \right)} +
{\hat J}_i^{\left( r+1 \right)}\right)
d {\mathbf{c}}_{i}=0 
\qquad {\left( l=1,2,3\right)} .
\label{system_r}
\end{eqnarray}

In (\ref{gas_dyn_system1}) and (\ref{system_r}), to simplify the further 
transformations according to the definitions of the pressure tensor of the 
$i$-th component
\begin{eqnarray}
{\mathrm{p}}_{i} \!\!&{\overset{\mathrm{def}}{=}}&\!\!
\int 
m_{i} \left( {\mathbf{c}}_{i}-{\mathbf{u}}_{i}\right)
\left( {\mathbf{c}}_{i}-{\mathbf{u}}_{i}\right)
f_{i}\,d {\mathbf{c}}_{i} \nonumber\\
\!\!&=&\!\! \int 
m_{i} {\mathbf{c}}_{i} {\mathbf{c}}_{i}
f_{i}\,d {\mathbf{c}}_{i} -
n_{i} m_{i} {\mathbf{u}}_{i} {\mathbf{u}}_{i} ,
\label{p_i}
\end{eqnarray}
the vector of heat flow density of the $i$-th component 
\begin{eqnarray}
{\mathbf{q}}_{i} \!\!&{\overset{\mathrm{def}}{=}}&\!\!
\int 
\frac{1}{2}\, m_{i} \left( {\mathbf{c}}_{i}-{\mathbf{u}}_{i}\right) ^{2}
\left( {\mathbf{c}}_{i}-{\mathbf{u}}_{i}\right)
f_{i}\,d {\mathbf{c}}_{i}\nonumber\\
\!\!&=&\!\! \int 
\frac{1}{2}\, m_{i} {c}_{i}^{2}
{\mathbf{c}}_{i} f_{i}\,d {\mathbf{c}}_{i}
- {\mathrm{p}}_{i}\cdot {\mathbf{u}}_{i}
- \frac{3}{2}\, n_{i} kT_{i}{\mathbf{u}}_{i}
\label{q_i}
\end{eqnarray}
and the temperature of the $i$-th component (\ref {condition3}), 
$\Psi_{i}^{\left( 1\right)}=m_i$, 
${\boldsymbol{\Psi}}_{i}^{\left( 2\right)}=m_{i}{\mathbf{C}}_{i}$, 
$\Psi_{i}^{\left( 3\right)}=\frac{1}{2}\, m_{i}C_{i}^{2}$, 
where ${\mathbf{C}}_{i}=\left( {\mathbf{c}}_{i}-{\mathbf{u}}_{i}\right)$, 
can be used instead of $\psi_{i}^{\left( l\right)}$. 

Functions $f_{i}^{\left( r\right)}$, $r=1,2\ldots $ are solutions to 
integral equations (\ref {sys_eq_series}), which can be rewritten as 
\begin{eqnarray}
{\mathcal D}_i^{\left( r \right)} +
\sum\limits_{j \ne i} J_{ij}^{\left( r \right)} +
{\hat J}_i^{\left( r \right)} = -{\check J}_i^{\left( r \right)} .
\label{r_eq}
\end{eqnarray}
The left-hand side of (\ref{r_eq}) includes only functions known from the 
previous step of the successive approximations method. 
Unknown function $f_{i}^{\left( r\right)}$ appears linearly only in the 
right-hand side of equation (\ref{r_eq}). 
Therefore, the general solution of equation (\ref{r_eq}) is 
$\Xi_{i}^{\left( r\right)} + \xi_{i}^{\left( r\right)}$, 
where $\xi_{i}^{\left( r\right)}$ is the 
general solution of homogeneous integral equation 
\begin{eqnarray}
{\check J}_i^{\left( r \right)} =
 J_i \left( {f_i^{\left( 0 \right)} f^{\left( {r} \right)} } \right) +
 J_i \left( {f_i^{\left( {r} \right)} f^{\left( 0 \right)} } \right)=0 ,
\label{r_eq_icr}
\end{eqnarray}
and $\Xi_{i}^{\left( r\right)}$ is some partial solution of 
inhomogeneous equation (\ref{r_eq}). 

Solutions $\Xi _{i}^{\left( r\right)}$ and $\xi _{i}^{\left( r\right)}$ will be 
sought in the form of 
$\Xi _{i}^{\left( r\right)} = f_{i}^{\left( 0\right)} 
\mathit{\Phi } _{i}^{\left( r\right)}$ and 
$\xi _{i}^{\left( r\right)} = f_{i}^{\left( 0\right)} 
\phi _{i}^{\left( r\right)}$, where 
$\mathit{\Phi } _{i}^{\left( r\right)}$ and $\phi _{i}^{\left( r\right)}$ 
are new unknown functions. 
In view of (\ref{res_H-equality}), 
\begin{eqnarray}
J_i \left( {f_i^{\left( 0 \right)}  
f^{\left( {0} \right)} \phi ^{\left( r\right)}} \right) &+&
J_i \left( f_i^{\left( {0} \right)}{\phi _{i}^{\left( r\right)} 
f^{\left( 0 \right)} } \right)
\nonumber\\*
&&=
\iint
f_i^{\left( 0 \right)} f^{\left( {0} \right)}
\left( 
\phi ^{\left( r\right)} + \phi _{i}^{\left( r\right)} -
\phi ^{\left( r\right) \prime } - \phi _{i}^{\left( r\right)\prime } 
\right)
k_{i} \,d{\mathbf{k}} \,d{\mathbf{c}} \nonumber\\*
&&{\overset{\mathrm{def}}{=}}\ n_{i}^{2}I_{i}\left( \phi ^{\left( r\right) }\right) =0 .
\label{r_eq_phi}
\end{eqnarray}
Having multiplied equation (\ref{r_eq_phi}) by 
$\phi _{i}^{\left( r\right)}\,d {\mathbf{c}}_{i}$, integrated over all 
${\mathbf{c}}_{i}$, and transformed the integrals with account 
for (\ref{equality}), we arrive at: 
\begin{eqnarray}
\frac{1}{4}\iiint
f_{i}^{\left( 0\right)} f^{\left( 0\right)}
\left(
\phi ^{\left( r\right)} + \phi _{i}^{\left( r\right)} -
\phi ^{\left( r\right) \prime } - \phi _{i}^{\left( r\right)\prime }
\right) ^{2}
k_{i} \,d {\mathbf{k}} \,d \mathbf{c}\,d {\mathbf{c}}_{i}=0 .
\label{phi-equality}
\end{eqnarray}
From (\ref {phi-equality}) we conclude, cf. with (\ref {H-equality}) and 
(\ref{ln_f_1}), that $\phi _{i}^{\left (r \right)}$ is a linear combination of 
additive collision invariants $\psi _{i}^{\left( l\right)}$: 
\begin{eqnarray}
\phi _{i}^{\left( r\right)}=
\alpha _{i}^{\left( 1,r\right) }\psi _{i}^{\left( 1\right)} +
\boldsymbol{\alpha }_{i}^{\left( 2,r\right) }{\boldsymbol{\psi}}_{i}^{\left( 2\right)} +
\alpha _{i}^{\left( 3,r\right) }\psi _{i}^{\left( 3\right)} ,
\label{phi_expr}
\end{eqnarray}
where $\alpha _{i}^{\left( 1,r\right)}$ and $\alpha _{i}^{\left( 3,r\right)}$ are 
arbitrary scalar functions of $\mathbf{r}$ and $t$, and 
$\boldsymbol{\alpha}_{i}^{\left( 2,r\right)}$ is an arbitrary vector function of 
$\mathbf{r}$ and $t$. 
In place of the additive invariants $\psi _{i}^{\left( l\right)}$, 
functions $\Psi _{i}^{\left( l\right)}$ can also be used, which are 
additive invariants of collision of particles of the same grade: 
\begin{eqnarray}
\phi _{i}^{\left( r\right)}=
\alpha _{i}^{\left( 1,r\right) \prime }\Psi _{i}^{\left( 1\right)} +
\boldsymbol{\alpha }_{i}^{\left( 2,r\right) \prime }
\cdot {\boldsymbol{\Psi}}_{i}^{\left( 2\right)} +
\alpha _{i}^{\left( 3,r\right) \prime }\Psi _{i}^{\left( 3\right)} ,
\label{phi_expr_Psi}
\end{eqnarray}
where $\alpha _{i}^{\left( 1,r\right) \prime}$ and 
$\alpha _{i}^{\left( 3,r\right) \prime}$ are new arbitrary scalar functions of 
$\mathbf{r}$ and $t$, and 
$\boldsymbol{\alpha}_{i}^{\left( 2,r\right) \prime}$ is a new arbitrary vector 
function of $\mathbf{r}$ and $t$. 
Thus, 
\begin{eqnarray}
\xi _{i}^{\left( r\right)}=
f_{i}^{\left( 0\right)}
{\left(
\alpha _{i}^{\left( 1,r\right)}+
\boldsymbol{\alpha }_{i}^{\left( 2,r\right)}\cdot m_{i}\mathbf{c}_{i}+
\alpha _{i}^{\left( 3,r\right)} \frac{1}{2}\,m_{i}c_{i}^{2} ,
\right)} .
\label{xi_expr}
\end{eqnarray}

To make use the results of the integral equation theory, transform equation 
(\ref{r_eq}) to the standard form. 
The right-hand side of integral equation (\ref{r_eq}), 
i.e. $n_{i}^{2}I_{i}\left( \mathit{\Phi } ^{\left( r\right)}\right)$, which is 
a function of $\mathbf{c}_{i}$ (and, naturally, of $\mathbf{r}$ and $t$, to simplify 
the notation, the evident dependencies are not specified) 
can be represented as 
\begin{eqnarray}
n_{i}^{2}I_{i}\left( \mathit{\Phi } ^{\left( r\right) }\right) &=&
K_{0}\left( \mathbf{c}_{i}\right)
\mathit{\Phi } _{i}^{\left( r\right) }\left( \mathbf{c}_{i}\right)
\nonumber\\*
&&+
\int
K\left( \mathbf{c}_{i},\mathbf{c}\right)
\mathit{\Phi } ^{\left( r\right) }\left( \mathbf{c}\right)
d \mathbf{c} ,
\label{right_eq_Phi}
\end{eqnarray}
where
\begin{eqnarray}
K_{0}\left( \mathbf{c}_{i}\right) &=&
\iint 
f_{i}^{\left( 0\right)} f^{\left( 0\right)} 
k_{i} \,d{\mathbf{k}} \,d\mathbf{c} 
\nonumber\\*
&=& f_{i}^{\left( 0\right)}\! \iiint 
f^{\left( 0\right)} 
\left\vert \mathbf{c}_{i}-\mathbf{c}\right\vert 
b\,db \,d\epsilon \,d{\mathbf{c}} ,
\label{K_0}
\end{eqnarray}
and $K\left( \mathbf{c}_{i}, \mathbf{c}\right)$ is a 
\textit{symmetric} function of $\mathbf{c}_{i}, \mathbf{c}$ 
(see \cite{hilbert12}, Chapter~XXII and \cite{chap60}, Chapter~7, \S ~6). 
Hence, equation (\ref {r_eq}) can be rewritten as: 
\begin{eqnarray}
F_{i}^{\left( r\right)}\left( \mathbf{c}_{i}\right) &=&
K_{0}\left( \mathbf{c}_{i}\right)
\mathit{\Phi } _{i}^{\left( r\right) }\left( \mathbf{c}_{i}\right)
\nonumber\\*
&&+
\int
K\left( \mathbf{c}_{i},\mathbf{c}\right)
\mathit{\Phi } ^{\left( r\right) }\left( \mathbf{c}\right)
d \mathbf{c} ,
\label{Phi_eq}
\end{eqnarray}
$F_{i}^{\left( r\right)}\left( \mathbf{c}_{i}\right)$ in (\ref{Phi_eq}) denotes the 
left-hand side of integral equation (\ref{r_eq}). 
Linear integral equation (\ref{Phi_eq}) is reduced by transformation
\begin{subequations}
 \label{Phi_trans}
 \begin{eqnarray}
  \mathit{\Phi } _{i}^{\left( r\right) }\left( \mathbf{c}_{i}\right) &=&
  \frac{\widetilde{ \mathit{\Phi } }_{i}^{\left( r\right)}\left( \mathbf{c}_{i}\right) }
  {\sqrt{K_{0}\left( \mathbf{c}_{i}\right) }},
  \label{Phi_transa} \\
  F_{i}^{\left( r\right)}\left( \mathbf{c}_{i}\right) &=&
  \widetilde{F}_{i}^{\left( r\right)}\left( \mathbf{c}_{i}\right)
  \sqrt{K_{0}\left( \mathbf{c}_{i}\right) } ,
  \label{Phi_transb} \\
  K\left( \mathbf{c}_{i},\mathbf{c}\right) &=&
  \widetilde{K}\left( \mathbf{c}_{i},\mathbf{c}\right)
  \sqrt{K_{0}\left( \mathbf{c}_{i}\right) K_{0}\left( \mathbf{c}\right) },
  \label{Phi_transc}
 \end{eqnarray}
\end{subequations}
maintaining the kernel symmetry to the \textit{linear integral equation of 
the second kind with symmetric kernel} 
\begin{eqnarray}
\widetilde{F}_{i}^{\left( r\right)}\left( \mathbf{c}_{i}\right) =
\widetilde{ \mathit{\Phi } }_{i}^{\left( r\right)}\left( \mathbf{c}_{i}\right)
+
\int
\widetilde{K}\left( \mathbf{c}_{i},\mathbf{c}\right)
\widetilde{ \mathit{\Phi } }^{\left( r\right)}\left( \mathbf{c}\right)
d \mathbf{c} .
\label{Phi_std_eq}
\end{eqnarray}
As homogeneous integral equation (\ref{r_eq_phi}) has nonzero solutions 
(\ref{phi_expr}), corresponding to equation (\ref{Phi_std_eq}), homogeneous 
integral equation 
\begin{eqnarray}
0 =
\tilde{ \phi }_{i}^{\,\left( r\right)}\left( \mathbf{c}_{i}\right)
+
\int
\widetilde{K}\left( \mathbf{c}_{i},\mathbf{c}\right)
\tilde{ \phi }^{\,\left( r\right)}\left( \mathbf{c}\right)
d \mathbf{c} .
\label{Phi_std_eq0}
\end{eqnarray}
has nonzero solutions 
\begin{eqnarray}
\tilde{ \phi }_{i}^{\,\left( r\right)}{\left( \mathbf{c}_{i}\right)} 
&=&
\phi _{i}^{\left( r\right)}{\left( \mathbf{c}_{i}\right)} 
{\sqrt{K_{0}\left( \mathbf{c}_{i}\right) }} \nonumber \\*
&=&
\left( 
\alpha _{i}^{\left( 1,r\right) }\psi _{i}^{\left( 1\right)} +
\boldsymbol{\alpha }_{i}^{\left( 2,r\right) }\cdot \boldsymbol{\psi }_{i}^{\left( 2\right)} +
\alpha _{i}^{\left( 3,r\right) }\psi _{i}^{\left( 3\right)}
\right)  
\nonumber \\*
&&\times 
{\sqrt{K_{0}\left( \mathbf{c}_{i}\right) }} .
\label{phi_w_expr}
\end{eqnarray}
Therefore, according to the second Fredholm alternative 
(\cite{courant89}, Chapter~III, \S \S ~2,~3 or 
\cite{korn68}, Section 15.3-7), 
if $\widetilde{K}\left( \mathbf{c}_{i}, \mathbf{c}\right)$ is piecewise-continuous and 
normalizable and 
$\widetilde{F}_{i}^{\left( r\right)}\left( \mathbf{c}_{i}\right)$ 
is continuous and square-integrable (these conditions are assumed met), 
the necessary and sufficient condition of solution existence of equation 
(\ref{Phi_std_eq}), with taking into account symmetry of kernel 
$\widetilde{K}\left( \mathbf{c}_{i}, \mathbf{c}\right)$, is the orthogonality of 
$ \widetilde{F}_{i}^{\left( r\right)}\left( \mathbf{c}_{i}\right)$ 
(with weight function 1) to each solution 
$\tilde{\phi}_{i}^{\,\left( r\right)}{\left( \mathbf{c}_{i}\right)}$ 
of equation (\ref{Phi_std_eq0}). 

Thus, the necessary and sufficient condition of the existence of a solution 
to equation (\ref{r_eq}) is the orthogonality of the left-hand side of the 
equation to functions 
$\psi _{i}^{\left( l\right)}\left(\mathbf{c}_{i}\right)$ 
[or, which is the same, to functions 
$ \Psi _{i}^{\left( l\right)}\left( \mathbf{c}_{i}\right)$], 
i.e. satisfaction of equalities (\ref{gas_dyn_system1}), (\ref{system_r}), 
which, on the other hand, can be considered as equations for determination of 
$n_{i}^{\left( r\right)}$, ${\mathbf{u}}_{i}^{\left( r\right)}$, 
$T_{i}^{\left( r\right)}$, $r=0,1,2\ldots$. 

The partial solution of inhomogeneous integral equation (\ref{r_eq}) 
$\Xi _{i}^{\left( r\right)} = f_{i}^{\left( 0\right)} 
\mathit{\Phi } _{i}^{\left( r\right)}$ can be constructed, for example, using expansion 
$\mathit{\Phi } _ {i} {\left (r \right)} \left (\mathbf {c} _ {i} \right) $ in series in 
terms of Sonin polynomials with expansion coefficients, depending on 
$\mathbf{r}$ and $t$, as 
this is done in \cite{chap60}, Chapter~7 and 8.

\section{Some remarks}
\label{sec:remarks}

By the successive approximations method we receive, generally speaking,
\textit{asymptotic} solution of the task.
Upper statement of the method for the kinetic Boltzmann equation solution 
is close to \cite{chap60}, Chapter~7 and 8. 
Somewhat more accurately the method for the kinetic Boltzmann equation solution
can be described within the theory of 
\textit{asymptotic expansions with variable coefficients}
\cite{bourbaki2004}, Chapter~V, \S ~2, Section~5.
Additional conditions (for example, initial or boundary conditions for a
differential equation), if there are ones, it is necessary to expand into 
asymptotic series, 
and to solve, equating coefficients at same terms of 
\textit{unified scale of comparison}
-- \cite{bourbaki2004}, Chapter~V, \S ~2, Section~1 
(usually, the scale of comparison is the set of functions $\theta ^r$, defined
on the neighborhood filter of the point $\theta =0$),
the resulting system of equations. 
If this system of equations can be resolved, we have asymptotic solution of the 
starting equation, satisfying supplementary conditions. 
Sometimes the found asymptotic solution is \textit{regular}
(\cite{lomov81}, Chapter~1), i.e. analytically depending on $\theta $, 
solution of the task.
For example, the equation 
(see \cite{cercignani75}, Chapter~V, \S ~2 or \cite{resibois77}, Chapter~IV, \S ~7.1)
\begin{eqnarray}
\theta \,\frac{df}{dt}+f=0 ,\qquad
f\left( 0\right) =0
\label{example0}
\end{eqnarray}
with zero initial condition have with regard to comparison scale 
of function $\theta ^r$, asymptotic solution 
$f^{\left( r\right)}\left( t\right) \equiv 0$, $\left( r=0,1,2\ldots \right)$, 
that is also the exact solution of the task.
However, the task
\begin{eqnarray}
\theta \,\frac{df}{dt}+f=0 ,\qquad
f\left( 0\right) =1
\label{example1}
\end{eqnarray}
has not analogous asymptotic solution, 
as the zero-order asymptotic solution of the differential equation 
$f^{\left( 0\right)}\left( t\right) \equiv 0$ 
contradicts to the initial condition $f^{\left( 0\right)}\left( 0\right) =1$.
This is not serious weakness of the successive approximations method.
In the task (\ref{example1}) 
one can introduce new function $g\left( t\right)$:
\begin{eqnarray}
f\left( t\right) = \exp {\left( {-t/\theta }\right)}\, g\left( t\right)
\label{new_function}
\end{eqnarray}
(cf. with \cite{lomov81}; 
though discussed by Lomov expansions are not asymptotic expansions 
with variable coefficients, and accordingly Lomov's approach
as a whole seems not quite correct, 
in the monography \cite{lomov81} is actually shown, that similar
replacements of functions allow to obtain asymptotic solutions for a
wide class of problems) and by the successive approximations method
receive asymptotic solution of the task 
$g^{\left( 0\right)}\left( t\right) \equiv \mathrm{const} =1$, 
$g^{\left( r\right)}\left( t\right) \equiv \mathrm{const} =0$ 
$\left( r=1,2\ldots \right)$, that is again the exact solution.

The criticism of the successive approximations method 
\cite{cercignani75}, Chapter~V, \S ~2 or \cite{resibois77},
Chapter~IV, \S ~7.1, possibly, reflects a dissatisfaction of authors 
with the unreasonable expansion of partial time derivative in the Enskog method.

Hilbert, having marked in \cite{hilbert12}, Chapter~XXII, that the
expansion
\begin{eqnarray}
F  = \frac{\Phi }
{\lambda } + {\Psi }  + X \lambda +\dotsb ,
\label{f_seriesHilbert}
\end{eqnarray}
analogous to (\ref{f_seriesEnscog}) upper, 
(Hilbert considered only one-component gas;
we maintain here Hilbert notation, however clear enough from a context)
is the power series in (small parameter) $\lambda $, 
satisfying to the Boltzmann equation and such, that expressions 
[cf. with 
(\ref{condition1})-(\ref{condition3n}) and
(\ref{n_series})-(\ref{T_series})]
\begin{eqnarray}
\int \psi ^{\left( i\right)} F\,d {\omega }
&=&\frac{1}{\lambda }
\int \psi ^{\left( i\right)} {\Phi }\,d {\omega }
+
\int \psi ^{\left( i\right)} {\Psi }\,d {\omega }
\nonumber\\
&&+ \,\lambda 
\int \psi ^{\left( i\right)} X\,d {\omega }
+\dotsb 
\qquad {\left( i=1,2,3,4,5\right)} 
\label{fint_seriesHilbert}
\end{eqnarray}
for $t=t_0$ pass into power series
\begin{eqnarray}
\Lambda ^{\left( i \right)}  = \frac{{f^{\left( i \right)} }}
{\lambda } + g^{\left( i \right)}  + \lambda h^{\left( i \right)} 
+\dotsb 
\qquad {\left( i=1,2,3,4,5\right)} ,
\label{f0int_seriesHilbert}
\end{eqnarray}
in the theorem, closing his work, has formulated 
"recipe" for (asymptotic) solution of the kinetic Boltzmann equation,
in which he has proposed five arbitrary functional parameters of functions 
${\Phi }, {\Psi }, X\ldots $ to define 
"\textit{from five partial differential equations}",
analogous (\ref{gas_dyn_system1}), (\ref{system_r}), 
"\textit{at that for $t=t_0$}" to preset
\begin{eqnarray}
\int \psi ^{\left( i\right)} {\Phi }\,d {\omega }
&=&{\lambda }\Lambda ^{\left( i \right)} 
\qquad {\left( i=1,2,3,4,5\right)} ,
\label{Phi_conditionHilbert}
\\
\int \psi ^{\left( i\right)} \Psi \,d {\omega }
&=&0 
\!\!\qquad \qquad {\left( i=1,2,3,4,5\right)} ,
\label{Psi_conditionHilbert}
\\
\int \psi ^{\left( i\right)} X\,d {\omega }
&=&0 
\!\!\qquad \qquad {\left( i=1,2,3,4,5\right)} .
\label{X_conditionHilbert}
\end{eqnarray}
In notation from (\ref{condition1})-(\ref{condition3n}), 
(\ref{n_series})-(\ref{T_series}), (\ref{gas_dyn_system1}),
(\ref{system_r}) upper, Hilbert proposed simply to use
special \textit{initial condition} 
\begin{eqnarray}
n\left( {\mathbf{r}},t_0\right) 
&=&
n^{\left( 0\right)}\left( {\mathbf{r}},t_0\right),
\label{n0_Hilbert} \\
{\mathbf{u}} \left( {\mathbf{r}},t_0\right) 
&=& 
{\mathbf{u}}^{\left( 0\right)}\left( {\mathbf{r}},t_0\right) ,
\label{u0_Hilbert} \\
T\left( {\mathbf{r}},t_0\right)  
&=& 
T^{\left( 0\right)}\left( {\mathbf{r}},t_0\right) 
\label{T0_Hilbert}
\end{eqnarray}
or
\begin{eqnarray}
\left. {\int \psi ^{\left( l\right)}
f^{\left( r\right)}\,d {\mathbf{c}}} \right|_{t = t_0 } 
\ {\overset{\mathbf{r}}{\equiv}}\ 0 
\qquad {\left( l=1,2,3\right)} 
\label{psi0_r}
\end{eqnarray}
for $r=1,2\ldots $.
It allowed him, as corollary of the theorem, to formulate
"fundamental result for the theory of gases: 
\textit {the state of stable gas at any $t$ is uniquely determinated, 
if for it at $t=t_0$ density, temperature and velocity are known as function
of a point of space}".

"For the further substantiation of the gas theory" it would be
necessary to supplement Hilbert's theorem with explicit definition
of five arbitrary functional parameters of functions 
$f _ {i} ^ {\left( r\right)} $, found on the $r$-th step
($r=0,1,2\ldots $) 
of the successive approximations method, 
through gas physical parameters (\ref {condition1}),
(\ref {condition2}), (\ref {condition3n}), (\ref {r_condition1}) -
(\ref {r_condition3n}), but Hilbert had not made it.

Enskog formulated up the Hilbert "recipe" for concrete calculations.
However meantime Enskog had made a logical mistake.
He used "null" conditions (\ref{psi0_r}) \textit{identically},
at any $t $, not just at $t=t_0$
(see \cite{chap60}, Chapter~7, \S ~1, Section~1):
\begin{eqnarray}
{\int \psi ^{\left( l\right)}
f^{\left( r\right)}\,d {\mathbf{c}}}
\ {\overset{\mathbf{r}{,}\,t}{\equiv}}\ 0 
\qquad {\left( l=1,2,3\right)} 
\label{psi_r_Enskog}
\end{eqnarray}
for $r=1,2\ldots $.
From the point of the successive approximations method view Enskog
instead of (\ref{n_series})-(\ref{T_series}) had supposed
\begin{eqnarray}
n\left( {\mathbf{r}},t,\theta \right) 
&=&
{\theta ^{\,0}}\,n\left( {\mathbf{r}},t,\theta \right) 
+{\theta ^1}\,0+{\theta ^2}\,0 +\dotsb ,
\label{n_Enskog} \\
{\mathbf{u}} \left( {\mathbf{r}},t,\theta \right) 
&=& 
{\theta ^{\,0}}\,{\mathbf{u}} \left( {\mathbf{r}},t,\theta \right) 
+{\theta ^1}\,0+{\theta ^2}\,0 +\dotsb ,
\label{u_Enskog} \\
T\left( {\mathbf{r}},t,\theta \right) 
&=&
{\theta ^{\,0}}\,T\left( {\mathbf{r}},t,\theta \right) 
+{\theta ^1}\,0+{\theta ^2}\,0 +\dotsb .
\label{T_Enskog}
\end{eqnarray}
If $n$, ${\mathbf{u}}$ and $T$ did not depend from $\mathbf{r}$ 
and $t$, it would mean, that Enskog used simultaneously different scales of
comparison
$\left \{ n\left( \theta \right) ,\theta ^1,\theta ^2 \ldots \right \}$,
$\left \{ {\mathbf{u}}\left( \theta \right) ,\theta ^1,\theta ^2
\ldots \right \}$,
$\left \{ T\left( \theta \right) ,\theta ^1,\theta ^2 
\ldots \right \}$ 
in the successive approximations method, that is already wrong.
In a general case, when $n$, ${\mathbf {u}}$ and $T$ depend from 
$\mathbf{r}$ and $t$, the sums (\ref{n_Enskog})-(\ref{T_Enskog}) 
cannot even be considered as asymptotic expansions with variable coefficients.

Infringement of logic of the successive approximations method
is immediately appeared in that from the equations, analogous 
(\ref{system_r}) ${\left( r=1,2,\ldots \right)}$,
in compliance with (\ref{psi_r_Enskog}) partial time derivatives vanish
\begin{eqnarray}
\int \psi ^{\left( l\right)} 
\frac{\partial f^{\left( r\right)}}{\partial t}d {\mathbf{c}}
=
\frac{\partial }{\partial t}
\int \psi ^{\left( l\right)} f^{\left( r\right)}d {\mathbf{c}}
=0
\qquad {\left( l=1,2,3\right)} ,
\label{timeD}
\end{eqnarray}
and with them terms of gas-dynamic equations, corresponding viscosity, heat 
conductivity \ldots , vanish.
Somehow to correct the situation, Enskog has been forced to enter
unreasonable expansion of partial time derivative (\ref{t_series_Enskog}).

\section{Calculation of definite multidimensional integrals}
\label{sec:integration}

In this section we are dealing with calculation of definite multidimensional 
integrals 
\begin{eqnarray}
\iiiint
\Psi _{i}^{\left( l\right)}\left( f_{i}^{\prime \left( 0\right)}
f_{j}^{\prime \left( 0\right)}-f_{i}^{\left( 0\right)}
f_{j}^{\left( 0\right)} \right)
g_{ij} b\,d b \,d \epsilon \,d {\mathbf{c}}_{i} \,d {\mathbf{c}}_{j} .
\label{int_common}
\end{eqnarray}
In (\ref{int_common}) 
$\Psi _{i}^{\left( 1\right)}= m_i$, 
${\boldsymbol{\Psi}}_{i}^{\left( 2\right)}= m_{i}{\mathbf{C}}_{i}$, 
$\Psi _{i}^{\left( 3\right)}= \frac{1}{2}\, m_{i} C_{i}^{2}$, 
${\mathbf{C}}_{i}={\mathbf{c}}_{i}-{\mathbf{u}}_{i}$;
\begin{eqnarray}
f_{i}^{\left( 0\right)}=
n_{i}\left( \frac{m_{i}}{2\pi kT_{i}}\right) ^{\frac{3}{2}}
e^{-{\left. m_{i}\left( {\mathbf{c}}_{i}-{\mathbf{u}}_{i}\right) ^{2} \right/ 
{2kT_{i}}}} ,
\label{maxwelldist}
\end{eqnarray}
is the Maxwell function of distribution of velocities of the $i$-th 
component particles, the prime in the distribution function means, that the 
distribution of the particle velocities ${\mathbf{c}}^{\prime}_{i}$ after the 
collision is considered. The other notation is specified above. 

According to (\ref{equality}), integral (\ref{int_common}) can be transformed 
as follows:
\begin{eqnarray}
&&\iiiint 
\Psi _{i}^{\left( l\right)}\left( f_{i}^{\prime \left( 0\right)}
f_{j}^{\prime \left( 0\right)}-f_{i}^{\left( 0\right)}
f_{j}^{\left( 0\right)} \right)
g_{ij} b\,db \,d\epsilon \,d{\mathbf{c}}_{i} \,d{\mathbf{c}}_{j} \nonumber\\*
&&\quad =
\iiiint
\Psi _{i}^{\left( l\right)}f_{i}^{\prime \left( 0\right)}
f_{j}^{\prime \left( 0\right)}
g_{ij}^{\prime } b^{\prime }\,db^{\prime } \,d\epsilon ^{\prime }\,
d{\mathbf{c}}^{\prime }_{i} \,d{\mathbf{c}}^{\prime }_{j} \nonumber\\*
&&\qquad -
\iiiint
\Psi _{i}^{\left( l\right)}f_{i}^{\left( 0\right)}
f_{j}^{\left( 0\right)}
g_{ij} b\,db \,d\epsilon \,d{\mathbf{c}}_{i} \,d{\mathbf{c}}_{j} \nonumber\\*
&&\quad =
\iiiint
\left(\Psi _{i}^{\left( l\right) \prime }-\Psi _{i}^{\left( l\right)} \right)
f_{i}^{\left( 0\right)}f_{j}^{\left( 0\right)}
g_{ij} b\,db \,d\epsilon \,d{\mathbf{c}}_{i} \,d{\mathbf{c}}_{j} .
\label{int_trans}
\end{eqnarray}

As the particle mass is conserved in the collision, for 
$\Psi _{i}^{\left( 1\right)} = m_i$ integral (\ref{int_trans}) vanishes. 
In the two other instances, 
generally speaking, this is not the case because there is no summation over 
the components, cf. with \cite{hirsch61}, Chapter~7, (2.33). 

Hereafter statements of the two following simple propositions are used 
several times. 

\begin{prop}
$f$ is assumed to be a ruled function on ${\mathbf{R}}$ with values in ${\mathbf{R}}$, 
${\mathbf{w}} \in {\mathbf{R}}^{3}$ be a fixed nonzero vector, 
${\mathbf {n}} \in {\mathbf{R}}^{3}$ be a unit vector. 
In this case
\begin{eqnarray}
\int\limits_{\Omega _{\mathbf{n}}}f\left( {\mathbf{w}}\cdot {\mathbf{n}}\right) {\mathbf{n}}
\,d \Omega _{\mathbf{n}}
=
\frac{2\pi \mathbf{w}}{w}\int\limits_{0}^{\pi }f\left( w\cos \left( \theta \right)
\right) \cos \left( \theta \right) \sin \left( \theta \right) d \theta .
\label{prop1_1}
\end{eqnarray}
In the left-hand side of (\ref{prop1_1}) the integral is taken over all 
directions of vector ${\mathbf{n}}$, 
${\mathbf{w}}\cdot {\mathbf{n}}$ is the scalar product of vectors 
${\mathbf{w}}$ and ${\mathbf{n}}$. 
\label{prop1}
\end{prop}

\begin{rem}
If ${\mathbf{w}}$ is a zero vector, then the right-hand side of (\ref {prop1_1}) is 
set equal to 0.
\end{rem}

\begin{proof}
Select the system of spherical coordinates, such that the 
polar axis direction be the same as the direction of the vector ${\mathbf{w}}$. 
Resolve the vector ${\mathbf{n}}$ into two components: 
parallel (${\mathbf{n}}_{\parallel}$) and perpendicular (${\mathbf{n}}_{\perp}$) 
to the vector ${\mathbf{w}}$:
\begin{eqnarray}
{\mathbf{n}}={\mathbf{n}}_{\parallel }+{\mathbf{n}}_{\perp }=
\frac{\left( {\mathbf{w}}\cdot {\mathbf{n}}\right) {\mathbf{w}}}{w^{2}}+
{\mathbf{n}}_{\perp } .
\label{prop1_2}
\end{eqnarray}
Having substituted expression (\ref{prop1_2}) for the vector ${\mathbf{n}} $ 
into the left-hand side of (\ref{prop1_1}) and integrated over the azimuthal 
angle, we obtain the required equality (\ref{prop1_2}), as in the integration 
over the azimuthal angle the ${\mathbf {n}}_{\perp}$ 
containing term vanishes.
\end{proof}

\begin{prop}
$E$ and $F$ is assumed to be two complete normalized spaces 
over field ${\mathbf{R}}$, 
${\mathbf {u}}$ be a continuous linear map of $E$ into $F$. 
In this case,
if ${\mathbf{f}}$ is a ruled function on interval $I\subset {\mathbf{R}}$ with its 
values in $E$, then ${\mathbf{u}\circ f}$ is the ruled function on $I$ with its 
values in $F$ and
\begin{eqnarray}
\int\limits_{a}^{b}{\mathbf{u}}\left( {\mathbf{f}}\left( t\right) \right) d t=
{\mathbf{u}}\left( \int\limits_{a}^{b}{\mathbf{f}}\left( t\right) d t\right) .
\label{prop2_1}
\end{eqnarray}
\label{prop2}
\end{prop}

\begin{proof}
Equality (\ref{prop2_1}) follows immediately from the expression for 
the derivative of composite function ${\mathbf{u}\circ f}$; 
the details of the proof 
can be found in \cite{bourbaki2004}, Chapter~II, \S ~1, Section~5.
\end{proof}

In these propositions ruled functions can be 
replaced by better known continuous functions.

The major difficulties in the calculation of integral (\ref{int_trans}) are 
associated with the fact that parameters of the Maxwell functions for the 
$i$-th and the $j$-th components are not equal: 
\begin{eqnarray}
{\mathbf{u}}_{i}\neq {\mathbf{u}}_{j},\qquad T_{i}\neq T_{j} .
\label{nequality}
\end{eqnarray}
As a result, it is not easy get rid of the scalar products 
of vectors in the exponent (it is desirable that the expression for the 
exponent be as simple as possible). 

As the scattering angle depends on the module of relative velocity of 
colliding particles [see, for example, \cite{chap60}, Chapter~3, \S ~4, Section~2 or 
\cite{hirsch61}, Chapter~1, (5.26)], it is natural to transfer in 
(\ref{int_trans}) to new 
variables -- center-of-mass velocity ${\mathbf{G}}_{ij}$ and 
relative colliding particle velocity ${\mathbf{g}}_{ij}$, which are related with the 
particle velocities ${\mathbf{c}}_{i}$ and ${\mathbf{c}}_{j}$ as: 
\begin{eqnarray}
{\mathbf{c}}_{i}&=&{\mathbf{G}}_{ij}+\frac{m_{j}}{m_{i}+m_{j}}\,{\mathbf{g}}_{ij} ,
\label{ciGg}
\\
{\mathbf{c}}_{j}&=&{\mathbf{G}}_{ij}-\frac{m_{i}}{m_{i}+m_{j}}\,{\mathbf{g}}_{ij} ,
\label{cjGg}
\end{eqnarray}
-- cf. with \cite{chap60}, Chapter~9, \S ~2.
For further simplification of the exponent vector 
${\mathbf{G}}_{ij}$ can be replaced by vector $\widetilde{\mathbf{G}}_{ij}$ 
resulting from $ {\mathbf{G}}_{ij}$ in an arbitrary affine transformation, for 
example, 
the one, which is a composition of shift, homothety (multiplication by a 
scalar), and rotation. The rotation arbitrariness is reduced to the freedom 
in choosing of direction of the polar axis in the transition to the 
spherical coordinate system. Similarly, the vector ${\mathbf{g}}_{ij}$ can be 
replaced by the vector $\tilde{\mathbf{g}}_{ij}$, resulting from 
${\mathbf{g}}_{ij}$ in composition of arbitrary homothety and arbitrary rotation. 
The shift of the origin of the vector ${\mathbf{g}}_{ij}$ would lead to a parametric 
dependence of the final integral on \textit{vectors} ${\mathbf {u}}_{i}$ and 
${\mathbf{u}}_{j}$
(cf. with \cite{oraevskiy85}, Chapter~3), which is undesirable, 
as integral (\ref{int_trans}) is supposed to be reduced to 
Chapman-Cowling integral $\Omega _{ij}^{\left( {l,s} \right)} $ 
[see \cite{chap60}, Chapter~9, \S ~3, (3.29) and 
\cite{hirsch61}, Chapter~7, (4.34)]. 

In view of the aforesaid, make the following substitution of variables 
${\mathbf{G}}_{ij}$ and ${\mathbf{g}}_{ij}$: 
\begin{eqnarray}
{\mathbf{g}}_{ij}&=&z_{1}\,\tilde{\mathbf{g}}_{ij} ,
\label{trans_g}
\\
{\mathbf{G}}_{ij}&=&z_{2}\,\widetilde{\mathbf{G}}_{ij}+z_{3}\,\tilde{\mathbf{g}}_{ij}+
\frac{{\mathbf{u}}_{i}+{\mathbf{u}}_{j}}{2} .
\label{trans_G}
\end{eqnarray}
In (\ref{trans_g})-(\ref{trans_G}) the scalar factors 
$z_{1}$, $z_{2}$, and $z_{3}$ are 
selected from the condition that the coefficients of 
$\tilde{\mathbf{g}}_{ij}^{2}$ and $ \widetilde{\mathbf{G}}_{ij}^{2}$ 
in the exponent be equal to 1 and the 
coefficient of the scalar product 
$\tilde{\mathbf{g}}_{ij}\cdot \widetilde{\mathbf{G}}_{ij}$ 
be equal to 0 (compare to the method of variable separation): 
\begin{eqnarray}
z_{1}&=&\sqrt{\frac{2\left( m_{i}T_{j}+m_{j}T_{i}\right)}{m_{i}m_{j}}} ,
\label{z_1}
\\
z_{2}&=&\sqrt{\frac{2T_{i}T_{j}}{m_{i}T_{j}+m_{j}T_{i}}} ,
\label{z_2}
\\
z_{3}&=&\frac{2\left( T_{i}-T_{j}\right)}{m_{i}+m_{j}}
\sqrt{\frac{m_{i}m_{j}}{2\left( m_{i}T_{j}+m_{j}T_{i}\right)}} \,.
\label{z_3}
\end{eqnarray}
Analogous substitutions of variables can be used in more complicated situations,
for example, discussed in \cite{oraevskiy85}, Chapter~3.

With new variables the exponent can be written in the following form: 
\begin{eqnarray}
-\left[ \tilde{g}_{ij}^{2}+\widetilde{G}_{ij}^{2}+a_{0}\,w^{2}+
a_{1}\tilde{\mathbf{g}}_{ij}\cdot {\mathbf{w}} +
a_{2}\widetilde{\mathbf{G}}_{ij}\cdot {\mathbf{w}}\right] ,
\label{simple_exp}
\end{eqnarray}
where
\begin{eqnarray}
{\mathbf{w}}&=&\frac{{\mathbf{u}}_{i}-{\mathbf{u}}_{j}}{2} ,
\label{w}
\\
a_{0}&=&\frac{m_{i}}{2T_{i}}+\frac{m_{j}}{2T_{j}} ,
\label{a_0}
\\
a_{1}&=&-2\sqrt{\frac{2m_{i}m_{j}}{m_{i}T_{j}+m_{j}T_{i}}} ,
\label{a_1}
\\
a_{2}&=&\left( \frac{m_{j}}{T_{j}}-\frac{m_{i}}{T_{i}}\right)
\sqrt{\frac{2T_{i}T_{j}}{m_{i}T_{j}+m_{j}T_{i}}} .
\label{a_2}
\end{eqnarray}

It is easy to see that it will be impossible to get rid of the constant term 
in exponent (\ref{simple_exp}) and, hence, of the constant exponential factor, which will 
appear hereafter in all expressions containing integrals of form 
(\ref{int_common}), (\ref{int_trans}) 
using only the above-specified transformations of variables (without using 
the shift of the origin of vector ${\mathbf{g}}_{ij}$). Such factors are missing in 
\cite{strum74}, (8). 

Determine Jacobian of transformation of variables 
$\left( {\mathbf{c}}_{i},{\mathbf{c}}_{j}\right) \longrightarrow \left(  
\tilde{\mathbf{g}}_{ij},\widetilde{\mathbf{G}}_{ij}\right)$ 
[see (\ref{trans_g})-(\ref{trans_G})]:
\begin{eqnarray}
\frac{\partial \left( {\mathbf{c}}_{i},{\mathbf{c}}_{j}\right)}
{\partial \left( \tilde{\mathbf{g}}_{ij},\widetilde{\mathbf{G}}_{ij}\right)}
&=&
\frac{\partial \left( {\mathbf{c}}_{i},{\mathbf{c}}_{j}\right)}
{\partial \left( {\mathbf{g}}_{ij},{\mathbf{G}}_{ij}\right)}
\,\frac{\partial \left( {\mathbf{g}}_{ij},{\mathbf{G}}_{ij}\right)}
{\partial \left( \tilde{\mathbf{g}}_{ij},\widetilde{\mathbf{G}}_{ij}\right)}
=z_{1}^{3}\,z_{2}^{3}\,\frac{\partial \left( {\mathbf{c}}_{i},{\mathbf{c}}_{j}\right)}
{\partial \left( {\mathbf{g}}_{ij},{\mathbf{c}}_{j}+\frac{m_{i}}{m_{i}+m_{j}}
\,{\mathbf{g}}_{ij}\right)}
\nonumber\\*
&=&
z_{1}^{3}\,z_{2}^{3}\,\frac{\partial \left( {\mathbf{c}}_{i},{\mathbf{c}}_{j}\right)}
{\partial \left( {\mathbf{g}}_{ij},{\mathbf{c}}_{j}\right)}
=z_{1}^{3}\,z_{2}^{3}\,\frac{\partial \left( {\mathbf{c}}_{i},{\mathbf{c}}_{j}\right)}
{\partial \left( {\mathbf{c}}_{i}-{\mathbf{c}}_{j},{\mathbf{c}}_{j}\right)}
=z_{1}^{3}\,z_{2}^{3} .
\label{jacobian}
\end{eqnarray}

Now consider the case, where 
$\Psi_{i}^{\left( l\right)}={\boldsymbol{\Psi}}_{i}^{\left( 2\right)}
= m_{i}\left( {\mathbf{c}}_{i}-{\mathbf{u}}_{i}\right)$. 
In view of (\ref{simple_exp}), (\ref{jacobian}), (\ref{trans_g})-(\ref{trans_G})
and the equality, following from the definition of $\mathbf{k}$ upper,
\begin{eqnarray}
m_{i}\left( {\mathbf{c}}_{i}^{\prime }-{\mathbf{c}}_{i}\right)
=\frac{m_{i}m_{j}}{m_{i}+m_{j}}
\left( {\mathbf{g}}_{ij}^{\prime } - {\mathbf{g}}_{ij}\right)
= -2\,\frac{m_{i}m_{j}}{m_{i}+m_{j}}
\left( {\mathbf{g}}_{ij}\cdot {\mathbf{k}}\right) {\mathbf{k}} 
\end{eqnarray}
integral 
(\ref{int_trans}) can be rewritten as: 
\begin{eqnarray}
&&\iiint
m_{i}\left( {\mathbf{c}}_{i}^{\prime }-{\mathbf{c}}_{i}\right)
f_{i}^{\left( 0\right)}f_{j}^{\left( 0\right)}
g_{ij} b\,db \,d\epsilon \,d{\mathbf{c}}_{i} \,d{\mathbf{c}}_{j} \nonumber\\*
\!&&\quad =
-2\,\frac{m_{i}m_{j}}{m_{i}+m_{j}}\,z_{1}^{5}\,z_{2}^{3}\,
n_{i}\left( \frac{m_{i}}{2\pi kT_{i}}\right) ^{\frac{3}{2}}
n_{j}\left( \frac{m_{j}}{2\pi kT_{j}}\right) ^{\frac{3}{2}}
\iiint
\left( \tilde{\mathbf{g}}_{ij}\cdot {\mathbf{k}}\right) {\mathbf{k}} \nonumber\\*
\!&&\qquad \times
\exp \left( -\left[ \tilde{g}_{ij}^{2}+\widetilde{G}_{ij}^{2}+
a_{0}\,w^{2}
+a_{1}\tilde{\mathbf{g}}_{ij}\cdot {\mathbf{w}}
+a_{2}\widetilde{\mathbf{G}}_{ij}\cdot {\mathbf{w}} \right] \right) \nonumber\\*
\!&&\qquad \times
\tilde{g}_{ij} \,b
\,d\epsilon \,d\widetilde{\mathbf{G}}_{ij} \,d\tilde{\mathbf{g}}_{ij} \,db .
\label{int_impulse}
\end{eqnarray}
Integrating with respect to $\epsilon $ in (\ref{int_impulse}) 
(with fixed $\tilde{\mathbf{g}}_{ij}$ and $\widetilde{\mathbf{G}}_{ij}$), resolve 
vector ${\mathbf{k}}$ into two components: 
the ones parallel and perpendicular to vector $\tilde{\mathbf{g}}_{ij}$ -- 
cf. with the proof of Proposition \ref{prop1}: 
\begin{eqnarray}
\int \left( \tilde{\mathbf{g}}_{ij}\cdot {\mathbf{k}}\right) {\mathbf{k}}\,d \epsilon
=2\pi \cos ^{2}\left( \frac{\pi -\chi }{2}\right) \tilde{\mathbf{g}}_{ij}
=\pi \left( 1-\cos \chi \right) \tilde{\mathbf{g}}_{ij} .
\label{int_epsilon}
\end{eqnarray}
When integrating over $\widetilde{\mathbf{G}}_{ij}$ and 
directions of vector $\tilde{\mathbf{g}}_{ij}$, use 
Proposition \ref{prop1}. As a result we arrive at 
\begin{eqnarray}
\!&&\iiint
m_{i}\left( {\mathbf{c}}_{i}^{\prime }-{\mathbf{c}}_{i}\right)
f_{i}^{\left( 0\right)}f_{j}^{\left( 0\right)}
g_{ij} b\,db \,d\epsilon \,d{\mathbf{c}}_{i} \,d{\mathbf{c}}_{j} \nonumber\\*
\!&&\quad =
16n_{i}n_{j}\,\frac{m_{i}T_{j}+m_{j}T_{i}}{\left( m_{i}+m_{j}\right)}
\,\frac{\mathbf{w}}{w}\,\frac{\sqrt{\pi }}{\xi ^{2}}
\,\,e^{-\frac{2m_{i}m_{j}w^{2}}{m_{i}T_{j}+m_{j}T_{i}}} \nonumber\\*
\!&&\qquad \times
\iint e^{-\tilde{g}_{ij}^{2}}
\left[ \tilde{g}_{ij}\xi \cosh \left( \tilde{g}_{ij}\xi \right) -
\sinh \left( \tilde{g}_{ij}\xi \right) \right]
\tilde{g}_{ij}^{2}\left( 1-\cos \chi \right) b\,db\,d\tilde{g}_{ij} .
\label{int_impulse_result}
\end{eqnarray}
In (\ref{int_impulse_result})
\begin{eqnarray}
\xi =a_{1}w ,
\label{xi}
\end{eqnarray}
factor $a_{1}$ is determined by formula (\ref{a_1}). 
It is easy to check that the 
singularity at $\xi =0$, which is possible when $w=0$, is actually absent 
in the right-hand side. 
Expression (\ref{int_impulse_result}) differs from 
Sruminskii's expression, i.e. \cite{strum74}, (8).

The case, where 
$\Psi _{i}^{\left( l\right)}=\Psi _{i}^{\left( 3\right)}= 
\frac{1}{2}\,m_{i}\left( {\mathbf{c}}_{i}-{\mathbf{u}}_{i}\right) ^{2}$, 
differs from the just considered one in the factor 
of the exponent in the right-hand side of (\ref{int_impulse}). 
Transform difference 
$\Psi _{i}^{\left( l\right) \prime}-\Psi _{i}^{\left( l\right)}$ 
according to (\ref{trans_g}), (\ref{trans_G}) and 
\cite{chap60}, Chapter~3, (4.9) and taking into 
account that only the relative particle velocity direction changes during 
the collision ($g_{ij} = g_{ij}^{\prime}$): 
\begin{eqnarray}
\Psi _{i}^{\left( 3\right) \prime }
&-&\Psi _{i}^{\left( 3\right)}=
\frac{m_{i}}{2}\left[ \left( {\mathbf{c}}_{i}^{\prime }-
{\mathbf{u}}_{i}\right) ^{2}-\left( {\mathbf{c}}_{i}-{\mathbf{u}}_{i}\right) ^{2}\right]
\nonumber\\*
&&=
\frac{m_{i}}{2}\left( {\mathbf{c}}_{i}^{\prime }-{\mathbf{c}}_{i}\right)
\left( {\mathbf{c}}_{i}^{\prime }+{\mathbf{c}}_{i}-2{\mathbf{u}}_{i}\right) 
\nonumber\\*
&&=
\frac{m_{i}m_{j}}{m_{i}+m_{j}}\left(
\left\{ {\mathbf{g}}_{ij}^{\prime }-{\mathbf{g}}_{ij}\right\}
\cdot
\left\{ {\mathbf{G}}_{ij}-{\mathbf{u}}_{i}\right\} \right) \nonumber\\*
&&=
-2\,z_{1}\,\frac{m_{i}m_{j}}{m_{i}+m_{j}}\left( \tilde{\mathbf{g}}_{ij} \cdot
{\mathbf{k}}\right) 
\nonumber\\*
&&\quad \times 
\left( {\mathbf{k}}\cdot \left\{ z_{2}\,\widetilde{\mathbf{G}}_{ij}+
z_{3}\,\tilde{\mathbf{g}}_{ij}-\frac{{\mathbf{u}}_{i}-{\mathbf{u}}_{j}}{2}
\right\} \right) .
\label{diff_psi}
\end{eqnarray}

With respect to its arguments the scalar product is a bilinear continuous 
function, therefore Proposition \ref{prop2} can be applied. 
On the integration with 
respect to $\epsilon $, similarly to (\ref{int_epsilon}), we arrive at: 
\begin{eqnarray}
\!\!\!\!\!\!\!\! &&-2\,z_{1}\,\frac{m_{i}m_{j}}{m_{i}+m_{j}}
\int \left( \tilde{\mathbf{g}}_{ij} \cdot
{\mathbf{k}}\right) \left( {\mathbf{k}}\cdot \left\{ z_{2}\,\widetilde{\mathbf{G}}_{ij}+
z_{3}\,\tilde{\mathbf{g}}_{ij}-\frac{{\mathbf{u}}_{i}-{\mathbf{u}}_{j}}{2}
\right\} \right) \,d\epsilon \nonumber\\*
\!\!\!\!\!\!\!\! &&\quad =
-2\pi \,z_{1}\,\frac{m_{i}m_{j}}{m_{i}+m_{j}}\left( 1-\cos \chi \right)
\left( \tilde{\mathbf{g}}_{ij}\cdot \left\{ z_{2}\,\widetilde{\mathbf{G}}_{ij}+
z_{3}\,\tilde{\mathbf{g}}_{ij}-\frac{{\mathbf{u}}_{i}-{\mathbf{u}}_{j}}{2}\right\}
\right) .
\label{int_epsilon_energy}
\end{eqnarray}
Perform the integration over $\widetilde{\mathbf{G}}_{ij}$ and 
directions of vector $\tilde{\mathbf{g}}_{ij}$ using 
Proposition \ref{prop1}:
\begin{eqnarray}
\!&&\iiint
\frac{m_{i}}{2}\left[ \left( {\mathbf{c}}_{i}^{\prime }-
{\mathbf{u}}_{i}\right) ^{2}-\left( {\mathbf{c}}_{i}-{\mathbf{u}}_{i}\right) ^{2}\right]
f_{i}^{\left( 0\right)}f_{j}^{\left( 0\right)}
g_{ij} b\,db \,d\epsilon \,d{\mathbf{c}}_{i} \,d{\mathbf{c}}_{j} \nonumber\\*
\!&&\quad =
16n_{i}n_{j}\frac{\sqrt{\pi }}{\xi }
\,e^{-\frac{2m_{i}m_{j}w^{2}}{m_{i}T_{j}+m_{j}T_{i}}}
\iint e^{-\tilde{g}_{ij}^{2}} \nonumber\\*
\!&&\qquad \times
\left\{ -\,D_{1,\,ij}\frac{w}{\xi }\left[ \tilde{g}_{ij}\xi
\cosh \left( \tilde{g}_{ij}\xi \right) -
\sinh \left( \tilde{g}_{ij}\xi \right) \right] -\,
2D_{2,\,ij}\tilde{g}_{ij}^{2}\sinh \left( \tilde{g}_{ij}\xi \right)
\right\} \nonumber\\*
\!&&\qquad \times
\tilde{g}_{ij}^{2}\left( 1-\cos \chi \right) b\,db\,d\tilde{g}_{ij} .
\label{int_energy_result}
\end{eqnarray}
In (\ref{int_energy_result}):
\begin{eqnarray}
D_{1,\,ij}&=&\frac{2\,m_{j}T_{i}}{m_{i}+m_{j}} ,
\label{C_1}
\\
D_{2,\,ij}&=&\frac{m_{i}m_{j}\left( T_{i}-T_{j}\right)}{2\,
\left( m_{i}+m_{j}\right) ^{2}}
\,\sqrt{\frac{2\,T_{i}}{m_{i}}+\frac{2\,T_{j}}{m_{j}}} .
\label{C_2}
\end{eqnarray}
The other notations are the same as in (\ref {int_impulse_result}). 

It is interesting to note that for ${\mathbf{u}}_{i}={\mathbf{u}}_{j}$ integral 
(\ref{int_impulse_result}) and the 
first term in (\ref{int_energy_result}) vanish and the second term in 
(\ref{int_energy_result}) is: 
\begin{eqnarray}
\sim const\left( T_{j}-T_{i}\right) ,
\end{eqnarray}
that corresponds to energy transfer from the "hot" components to the 
"cold", see the gas-dynamic equations system below. 
In view of the sign of 
$a_{1}$ (\ref{a_1}) and definition of $\xi $ (\ref{xi}), 
the first term leads to \textit{temperature increase} with $w\neq 0$.

\section{First-order equations system of multi-component nonequilibrium gas-dynamics}
\label{sec:order1}

Above the gas-dynamic equations system has 
been derived, in a sense, as a "by-product" during the Boltzmann equation 
solution. 
More generally, the gas-dynamic equations system can be written in 
the form of the transport equations, 
cf. with \cite{chap60}, Chapter~3, (1.12) and 
\cite{hirsch61}, Chapter~7, (2.31). 

Having multiplied the Boltzmann equation for the $i$-th component 
(\ref{boltzmannend}) by 
$\Psi _{i}^{\left( l\right)}$ and integrated over all values of ${\mathbf{c}}_{i}$ 
(it is assumed that all the integrals obtained below converge and products 
like $\Psi _{i}^{\left( l\right)}{\mathbf{X}}_{i} f_{i}$ tend to zero, when 
${\mathbf{c}}_{i}$ tends to infinity), we arrive at: 
\begin{eqnarray}
\!&&\int \Psi _{i}^{\left( l\right)}\left( \frac{\partial f_{i}}{\partial t}+
{\mathbf{c}}_{i}\cdot \frac{\partial f_{i}}{\partial {\mathbf{r}}} +
\frac{{\mathbf{X}}_{i}}{m_{i}}\cdot \frac{\partial f_{i}}{\partial
{\mathbf{c}}_{i}}\right) d{\mathbf{c}}_{i} \nonumber\\*
\!&&\quad =
\sum\limits_{j}\iiint \Psi _{i}^{\left( l\right)}
\left( f_{i}^{\prime }f_{j}^{\prime }-f_{i}f_{j}\right)
g_{ij}b\,db\,d\epsilon \,d{\mathbf{c}}_{i}\,d{\mathbf{c}}_{j} .
\label{phi_boltzmann}
\end{eqnarray}
The terms in the left-hand side of this equation can be transformed: 
\begin{eqnarray}
\int \Psi _{i}^{\left( l\right)}\frac{\partial f_{i}}
{\partial t}\,d{\mathbf{c}}_{i}
&=&
\frac{\partial }{\partial t}\int \Psi _{i}^{\left( l\right)}f_{i}\,d{\mathbf{c}}_{i}-
\int \frac{\partial \Psi _{i}^{\left( l\right)}}{\partial t}f_{i}\,d{\mathbf{c}}_{i}
\nonumber\\*
&=&
\frac{\partial \left( n_{i}\overline{\Psi _{i}^{\left( l\right)}}\right)}{\partial t}-
n_{i}\overline{\frac{\partial \Psi _{i}^{\left( l\right)}}{\partial t}}\ ,
\label{trans1}
\\
\int \Psi _{i}^{\left( l\right)}{\mathbf{c}}_{i}\cdot 
\frac{\partial f_{i}}{\partial {\mathbf{r}}}\,d{\mathbf{c}}_{i}
&=&
\frac{\partial }{\partial {\mathbf{r}}}\cdot \int \Psi _{i}^{\left( l\right)}
{\mathbf{c}}_{i}f_{i}\,d{\mathbf{c}}_{i} 
-\int {\mathbf{c}}_{i}\cdot \frac{\partial \Psi _{i}^{\left( l\right)}}
{\partial {\mathbf{r}}}f_{i}\,d{\mathbf{c}}_{i} \nonumber\\*
&=&
\frac{\partial }{\partial {\mathbf{r}}}
\cdot n_{i}\overline{\Psi _{i}^{\left( l\right)}{\mathbf{c}}_{i}} 
-n_{i}\overline{{\mathbf{c}}_{i}
\cdot \frac{\partial \Psi _{i}^{\left( l\right)}}{\partial {\mathbf{r}}}}\ ,
\label{trans2}
\\
\int \Psi _{i}^{\left( l\right)}\frac{{\mathbf{X}}_{i}}{m_{i}}
\cdot \frac{\partial f_{i}}{\partial {\mathbf{c}}_{i}}
\,d{\mathbf{c}}_{i}
&=&
-\int \left( \frac{\partial }{\partial {\mathbf{c}}_{i}}
\cdot \Psi _{i}^{\left( l\right)}\frac{{\mathbf{X}}_{i}}{m_{i}}\right) f_{i}
\,d{\mathbf{c}}_{i} \nonumber\\*
&=&
-n_{i}\overline{\frac{\partial }{\partial {\mathbf{c}}_{i}}
\cdot \Psi _{i}^{\left( l\right)}\frac{{\mathbf{X}}_{i}}{m_{i}}}\ .
\label{trans3}
\end{eqnarray}
In (\ref{trans1})-(\ref{trans3}), the bar, as usually, denotes the average of 
the quantity
\begin{eqnarray}
\overline{V}_{i}=\frac{1}{n_{i}}\int V_{i} f_{i}\,d {\mathbf{c}}_{i} ;
\label{average}
\end{eqnarray}
${\mathbf{r}}$ and ${\mathbf{c}}_{i}$ are considered as independent variables. 
In view of (\ref{trans1})-(\ref{trans3}), from (\ref{phi_boltzmann}) we obtain
\begin{eqnarray}
\!&&\frac{\partial \left( n_{i}\overline{\Psi _{i}^{\left( l\right)}}\right)}
{\partial t}
+\frac{\partial }{\partial {\mathbf{r}}}
\cdot n_{i}\overline{\Psi _{i}^{\left( l\right)}{\mathbf{c}}_{i}} 
-n_{i}
\left( \overline{\frac{\partial \Psi _{i}^{\left( l\right)}}{\partial t}}
+\overline{{\mathbf{c}}_{i}\cdot \frac{\partial \Psi _{i}^{\left( l\right)}}
{\partial {\mathbf{r}}}}
+\overline{\frac{\partial }{\partial {\mathbf{c}}_{i}}
\cdot \Psi _{i}^{\left( l\right)}
\frac{{\mathbf{X}}_{i}}{m_{i}}}\right) \nonumber\\*
\!&&\quad =
\sum\limits_{j}\iiint \Psi _{i}^{\left( l\right)}
\left( f_{i}^{\prime }f_{j}^{\prime }-f_{i}f_{j}\right)
g_{ij}b\,db\,d\epsilon \,d{\mathbf{c}}_{i}\,d{\mathbf{c}}_{j} ,
\label{trans_eq}
\end{eqnarray}
-- the \textit{transfer equation} for the $\Psi _{i}^{\left( l\right)}$, 
that refers to particles of the $i$-th grade.

To derive the equations of mass, momentum and energy transport for the 
$i$-th component from (\ref {trans_eq}), sequentially substitute 
$m_i$, $m_{i}\left( {\mathbf{c}}_{i}-{\mathbf{u}}_{i}\right)$, 
$\frac{1}{2}\, m_{i}\left( {\mathbf{c}}_{i}-{\mathbf{u}}_{i}\right) ^{2}$  
for $\Psi _{i}{\left( l\right)}$ into (\ref{trans_eq}).

In the Enskog-Chapman theory, in view of the additional summation over $i$, 
the right-hand side of (\ref{trans_eq}) vanishes always. 
However, if the velocity distribution functions for some components are Maxwell 
functions (\ref{maxwelldist}) with 
different parameters of mean velocity and temperature 
(${\mathbf {u}}_{i} \neq {\mathbf{u}}_{j}$, $T_{i} \neq T_{j}$), for example, due to 
some external effects (see below), then nonzero terms remain in the right-hand 
side of (\ref{trans_eq}). 
In this case equations (\ref{trans_eq}) ${\left( l=1,2,3\right)}$ are
the same as equations (\ref{gas_dyn_system1}).
Thus, on straightforward transformations we arrive at the following 
gas-dynamic equations system 
[cf. with \cite{hirsch61}, Chapter~7, (2.42), (2.45), (2.47)]: 
\begin{eqnarray}
\frac{\partial n_{i}}{\partial t}&=&
-\frac{\partial }{\partial {\mathbf{r}}}
\cdot n_{i}{\mathbf{u}}_{i} ,
\label{trans_m}
\\
n_{i}m_{i}\frac{\partial {\mathbf{u}}_{i}}{\partial t}
+\frac{\partial }{\partial {\mathbf{r}}}
\cdot {\mathrm{p}}_{i}^{\,\left( 0\right)}
-\sum\limits_{j\neq i}{\mathbf{I}}_{p,\,ij}^{\,\left( 0\right)}
&=&
n_{i}{\mathbf{X}}_{i}-n_{i}m_{i}{\mathbf{u}}_{i}
\cdot \frac{\partial }{\partial {\mathbf{r}}}\ {\mathbf{u}}_{i} ,
\label{trans_p}
\\
\frac{\partial \hat{E}_{i}}{\partial t} 
+\frac{\partial }{\partial {\mathbf{r}}}\cdot {\mathbf{q}}_{i}^{\,\left( 0\right)} 
+{\mathrm{p}}_{i}^{\,\left( 0\right)}:\frac{\partial {\mathbf{u}}_{i}}
{\partial {\mathbf{r}}}
-\sum\limits_{j\neq i}I_{e,\,ij}^{\,\left( 0\right)}
&=& -\frac{\partial }{\partial {\mathbf{r}}}
\cdot \hat{E}_{i}{\mathbf{u}}_{i} .
\label{trans_e}
\end{eqnarray}
In (\ref{trans_m})-(\ref{trans_e}):
\begin{eqnarray}
{\mathrm{p}}_{i}^{\,\left( 0\right)}=n_{i}m_{i}\overline{\left( {\mathbf{c}}_{i}-{\mathbf{u}}_{i}\right)
\left( {\mathbf{c}}_{i}-{\mathbf{u}}_{i}\right)}^{\,\left( 0\right)}
=n_{i} k T_{i} {\mathrm{U}}=p_{i}^{\,\left( 0\right)} {\mathrm{U}}
\label{p_i_0}
\end{eqnarray}
is the $i$-th component pressure tensor, 
$p_{i}^{\,\left( 0\right)}$ is the hydrostatic pressure, 
${\mathrm{U}}$ is the unit tensor,
\textit{double product} of two second rank tensors ${\mathrm{w}}$ and 
${\mathrm{w}}^{\prime }$ 
(\cite{chap60}, Chapter~1, \S~3) is the scalar
${\mathrm{w}}:{\mathrm{w}}^{\prime }=
\sum_{\alpha }\sum_{\beta }w_{\alpha \beta }w^{\prime }_{\beta \alpha }=
{\mathrm{w}}^{\prime }:{\mathrm{w}}$,
\begin{eqnarray}
{\mathbf{q}}_{i}^{\,\left( 0\right)}=\frac{1}{2}\,n_{i}m_{i}\overline{\left( 
{\mathbf{c}}_{i}-
{\mathbf{u}}_{i}\right) ^{2}\left( {\mathbf{c}}_{i}-
{\mathbf{u}}_{i}\right)}^{\,\left( 0\right)} = 0
\label{q_i_0}
\end{eqnarray}
is the $i$-th component heat flux density vector,
\begin{eqnarray}
\hat{E}_{i}=
\frac{1}{2}\,n_{i}m_{i}\overline{\left( {\mathbf{c}}_{i}-
{\mathbf{u}}_{i}\right) ^{2}}^{\,\left( 0\right)}
\label{E_i_0}
\end{eqnarray}
is the internal energy of particles of the $i$-th component per unit volume, 
which is equal, in this case, to energy of their translational motion, 
however, the energy transfer equation, written in form (\ref{trans_e}), 
apparently, can be used in more general cases as well 
(cf. with \cite{hirsch61}, Chapter~7, \S ~6); 
in (\ref{p_i_0})-(\ref{E_i_0}) superscript $^{\left( 0\right)}$ denotes 
averaging (\ref{average}) with Maxwell function $f_{i}^{\left( 0\right)}$ from 
(\ref {maxwelldist}). 

In (\ref{trans_p})-(\ref{trans_e}) 
${\mathbf{I}}_{{\mathrm{p}},\,ij}^{\,\left( 0\right)}$, 
$I_{e,\,ij}^{\,\left( 0\right)}$ denote integrals (\ref{int_impulse_result}) 
and (\ref{int_energy_result}), respectively. 
When averaging the last term in the left-hand side of (\ref{trans_eq}), 
external force ${\mathbf{X}}_{i}$, acting on the particle of the $i$-th grade, is 
assumed independent of the particle velocity.

\section{Values of kinetic integrals for interaction potential of rigid spheres}
\label{sec:integrals}

Integral terms ${\mathbf{I}}_{{\mathrm{p}},\,ij}^{\,\left( 0\right)}$, 
$I_{e,\,ij}^{\,\left( 0\right)}$, appearing in multi-component gas-dynamics 
equations system (\ref{trans_m})-(\ref{trans_e}), are quite complex functions of 
mean velocities and temperatures of 
separate components, mainly, because of a complex dependence of deflection 
angle $\chi $ on relative velocity of colliding particles 
[cf. with \cite{hirsch61}, Chapter~1, (5.26)]:
\begin{eqnarray}
\chi \left( b,g\right) =
\pi -2b\int\limits_{r_{m}}^{\infty }\frac{d r/r^{2}}
{\sqrt{1-\frac{\varphi \left( r\right)}
{\frac{1}{2}\mu g^{2}}-\frac{b^{2}}{r^{2}}}} .
\label{chi2}
\end{eqnarray}
The component temperatures appear in the resultant expressions as a 
making non-dimensional factor. 

In the simplest case of particles, interacting according to the law of rigid 
spheres, the following analytical expressions for 
${\mathbf{I}}_{{\mathrm{p}},\,ij}^{\,\left( 0\right)}$, and 
$I_{e,\,ij}^{\,\left( 0\right)}$ have been derived: 
\begin{eqnarray}
{\mathbf{I}}_{p,\,ij}^{\,\left( 0\right)}&=&
  n_i n_j \,\frac{{m_i T_j  + m_j T_i }}
{{m_i  + m_j }}\,\frac{\mathbf{w}}
{w}\,\frac{{\sqrt \pi  }}
{{2\xi ^2 }}\ \sigma _{ij}^2 \nonumber\\*
&&\times 
\left[ {\exp \left( { - \frac{{\xi ^2 }}
{4}} \right)2\xi \left( {\xi ^2  + 2} \right) + 
\sqrt \pi  \left( {\xi ^4  + 4 \xi ^2  - 4} \right){\mathrm{erf}}
\left( {\frac{\xi }{2}} \right)} \right] ,
\label{I_p}
\\
I_{e,\,ij}^{\,\left( 0\right)}&=&
- n_i n_j \,\frac{{\sqrt \pi  }}
{{2\xi ^2 }} \,\sigma _{ij}^2
\exp \left( { - \frac{{\xi ^2 }}
{4}} \right)\left[ {2 D_{1,\,ij} w\xi \left( {\xi ^2  + 2} \right) +
2 D_{2,\,ij}\,\xi ^2 \left( {\xi ^2  + 10} \right)} \right] 
\nonumber\\*
&&- 
n_i n_j \,\frac{{\pi  }} {{2\xi ^2 }}\,\sigma _{ij}^2
\nonumber\\*
&&\times  
\left[ {D_{1,\,ij} w\left( {\xi ^4  + 4\xi ^2  - 4} \right) +
D_{2,\,ij} \,\xi \left( {\xi ^4  + 12\xi ^2  + 12} \right)} \right]
{\mathrm{erf}}\left( {\frac{\xi }
{2}} \right) .
\label{I_e}
\end{eqnarray}
In (\ref{I_p})-(\ref{I_e}) notations from (\ref{w}), (\ref{a_1}), (\ref{xi}), 
(\ref{C_1})-(\ref{C_2}) are used.

\section{Proposed method and Enskog-Chapman theory}
\label{sec:versus}

Let's consider, what do nonzero conditions
(\ref{r_condition1})-(\ref{r_condition3n}) and expansions
(\ref{n_series})-(\ref{T_series}) for one-component gas lead to;
for one-component gas the Enskog approach and the Struminskii approach
coincide.

Differential equations (\ref{gas_dyn_system1}), from which 
functions $n^{\left( 0\right)}(\mathbf{r},t)$, 
${\mathbf{u}}^{\left( 0\right)}(\mathbf{r},t)$, 
$T^{\left( 0\right)}(\mathbf{r},t)$ 
are found, for one-component gas can be written in the form 
[cf. with (\ref{trans_m})-(\ref{E_i_0})]:
\begin{eqnarray}
\frac{D n^{\left( 0\right)}}{D t}&=&
-n^{\left( 0\right)}\frac{\partial }{\partial {\mathbf{r}}}
\cdot {{\mathbf{u}}^{\left( 0\right)}} ,
\label{trans_m1}
\\
n^{\left( 0\right)}m\,\frac{D {{\mathbf{u}}^{\left( 0\right)}}}{D t}
&=& n^{\left( 0\right)}{\mathbf{X}}-
\frac{\partial p^{\left( 0\right)}}{\partial {\mathbf{r}}} ,
\label{trans_p1}
\\
\frac{D T^{\left( 0\right)}}{D t} 
= - \frac{2p^{\left( 0\right)}}{3n^{\left( 0\right)}k}\frac{\partial }
{\partial {\mathbf{r}}}
\cdot {{\mathbf{u}}^{\left( 0\right)}}
&=& - \frac{2T^{\left( 0\right)}}{3}\frac{\partial }
{\partial {\mathbf{r}}}
\cdot {{\mathbf{u}}^{\left( 0\right)}} .
\label{trans_e1}
\end{eqnarray}
In (\ref{trans_m1})-(\ref{trans_e1})
\begin{eqnarray}
\frac{D }{D t} 
&=& \frac{\partial }{\partial t}
+ {{\mathbf{u}}^{\left( 0\right)}} 
\cdot \frac{\partial }{\partial {\mathbf{r}}} ,
\label{def_Dt}
\\
p^{\left( 0\right)} &=& n^{\left( 0\right)} k T^{\left( 0\right)} .
\label{def_p}
\end{eqnarray}

From (\ref{trans_m1}), (\ref{trans_e1}) we receive, that in the first 
order gas-dynamic flow is \textit{adiabatic}:
\begin{eqnarray}
\frac{D }{D t} \left[ {n^{\left( 0\right)} 
\left( T^{\left( 0\right)}\right)^{-\frac{3}{2}}} \right]=0 .
\label{adiabatic}
\end{eqnarray}

For one-component gas integral equation (\ref{r_eq}), $r=1$, 
from which  
$f^{\left( 1\right)}=f^{\left( 0\right)}\mathit{\Phi } ^{\left( 1\right)}$
is found,
with taking into account (\ref{maxwell_f0}), (\ref{right_eq_Phi}) and
(\ref{trans_m1})-(\ref{adiabatic}),     
can be written as (cf. with \cite{chap60}, Chapter~7, \S ~3):
\begin{eqnarray}
- n^{2}I\left( \mathit{\Phi } ^{\left( 1\right) }\right) 
&=&
\frac{\partial f^{\left( 0\right)}}{\partial t}
+{\mathbf{c}}\cdot \frac{\partial f^{\left( 0\right)}}{\partial \mathbf{r}} 
+\frac{\mathbf{X}}{m}\cdot \frac{\partial f^{\left( 0\right)}}
{\partial {\mathbf{c}}} \nonumber\\*
&=&
f^{\left( 0\right)} 
\left[
\left( {\mathcal C}^{2}-\frac{5}{2}\right) \mathbf{C}
\cdot \frac{\partial \ln T^{\left( 0\right)}}{\partial \mathbf{r}} 
+2\,\overset{\circ }{{\boldsymbol{\mathcal C}}{\boldsymbol{\mathcal C}}}:
\frac{\partial }{\partial \mathbf{r}}\,\mathbf{u}^{\left( 0\right)}\right] .
\label{f1_eq}
\end{eqnarray}
In (\ref{f1_eq})
${\mathbf{C}} = {\mathbf{c}} - {\mathbf{u}}^{\left( 0\right)}$,
\begin{eqnarray}
\boldsymbol{\mathcal C}
= \left( \frac{m}{2kT^{\left( 0\right) }}\right) ^{\frac{1}{2}}\mathbf{C},
\label{mathcalC}
\end{eqnarray}
${\mathcal C}$ is the module of vector $\boldsymbol{\mathcal C}$; 
for arbitrary second-rank tensor ${\mathrm{w}}$ 
\begin{eqnarray}
\overset{\circ }{\mathrm{w}}={\mathrm{w}}-\frac{1}{3}\,{\mathrm{U}}
\left( {\mathrm{U}}:{\mathrm{w}}\right)
\label{wcirc}
\end{eqnarray}
-- tensor with zero trace.

The general (scalar) solution of equation (\ref{f1_eq}), being the sum of 
some partial solution of equation (\ref{f1_eq}) and  
general solution of homogeneous equation 
$I\left( \phi ^{\left( 1\right) }\right)=0$, cf. with (\ref{phi_expr_Psi}), 
we can seek in the form of:
\begin{eqnarray}
f^{\left( 1\right) }
&=&
f^{\left( 0\right) }
\left[ -\frac{1}{n^{\left( 0\right) }}\left( \frac{2kT^{\left( 0\right) }}{m}\right)^{\frac{1}{2}} 
\mathbf{A}\cdot \frac{\partial \ln T^{\left( 0\right) }}{\partial \mathbf{r}}
-\frac{1}{n^{\left( 0\right) }}\,\mathrm{B}:
\frac{\partial }{\partial \mathbf{r}}\mathbf{u}^{\left( 0\right)}\right] \nonumber\\*
&&+ \,
f^{\left( 0\right) }
\left( \alpha ^{\left( 1,1\right) }
+\boldsymbol{\alpha }^{\left( 2,1\right)}\cdot m\mathbf{C}
+\alpha ^{\left( 3,1\right) }\frac{1}{2}\,mC^{2}\right) ,
\label{f1_solution}
\end{eqnarray}
where the vector function $\mathbf{A}$ is partial solution of the equation
\begin{eqnarray}
nI\left( \mathbf{A}\right) 
=f^{\left( 0\right)} 
\left( {\mathcal C}^{2}-\frac{5}{2}\right) \boldsymbol{\mathcal C} ,
\label{A_eq}
\end{eqnarray}
and the tensor function $\mathrm{B}$ is partial solution of the equation
\begin{eqnarray}
nI\left( \mathrm{B}\right) 
=2f^{\left( 0\right)} 
\overset{\circ }{{\boldsymbol{\mathcal C}}{\boldsymbol{\mathcal C}}} .
\label{B_eq}
\end{eqnarray}

Solvability conditions of equations (\ref{A_eq}) and (\ref{B_eq}) are satisfied,
i.e. (see the section \ref{sec:method} upper):
\begin{eqnarray}
\int \Psi ^{\left( l\right)} 
f^{\left( 0\right)} 
\left( {\mathcal C}^{2}-\frac{5}{2}\right) \boldsymbol{\mathcal C}
\,d {\mathbf{c}}&=&0 
\qquad {\left( l=1,2,3\right)} ,
\label{A_exist}
\\
\int \Psi ^{\left( l\right)} 
f^{\left( 0\right)} 
\overset{\circ }{{\boldsymbol{\mathcal C}}{\boldsymbol{\mathcal C}}}
\,d {\mathbf{c}}&=&0 
\qquad {\left( l=1,2,3\right)} .
\label{B_exist}
\end{eqnarray}
In (\ref{A_exist}) and (\ref{B_exist})
$\Psi ^{\left( 1\right)}= m$,
${\boldsymbol{\Psi}}^{\left( 2\right)}= m{\mathbf{C}}$,
$\Psi ^{\left( 3\right)}= \frac{1}{2}\, m C^{2}$.

Because $\mathbf{r}$,~$t$ and $\mathbf{u}^{\left( 0\right)}$ 
do not explicitly occur in equations (\ref{A_eq}) and (\ref{B_eq}), 
and right-hand side of equation (\ref{B_eq}) is symmetric tensor with  
zero trace, solutions $\mathbf{A}$ and $\mathrm{B}$ can be 
sought in the form of:
\begin{eqnarray}
\mathbf{A}
&=&A\left( n^{\left( 0\right) }, {\mathcal C}, T^{\left( 0\right) }\right) 
\boldsymbol{\mathcal C} ,
\label{A_expr}
\\
\mathrm{B}
&=&B\left( n^{\left( 0\right) }, {\mathcal C}, T^{\left( 0\right) }\right) 
\overset{\circ }{{\boldsymbol{\mathcal C}}{\boldsymbol{\mathcal C}}} ,
\label{B_expr}
\end{eqnarray}
where $A\left( n^{\left( 0\right) }, {\mathcal C}, T^{\left( 0\right) }\right)$ and
$B\left( n^{\left( 0\right) }, {\mathcal C}, T^{\left( 0\right) }\right)$ --
scalar functions of $n^{\left( 0\right) }$,~${\mathcal C}$ and 
$T^{\left( 0\right) }$.
It is possible to impose an additional condition on the solution $\mathbf{A}$
(cf. with \cite{chap60}, Chapter~7, \S ~3, Section~1):
\begin{eqnarray}
\int {\mathcal C}^2 f^{\left( 0\right)}
A\left( n^{\left( 0\right) }, {\mathcal C}, T^{\left( 0\right) }\right)
\,d {\mathbf{c}}=0 .
\label{A_condition}
\end{eqnarray}

Having substituted (\ref{f1_solution}) in (\ref{r_condition1})-(\ref{r_condition3n}) 
with taking into account (\ref{maxwell_f0}), (\ref{param_f}) and
(\ref{A_condition}), we arrive at:
\begin{eqnarray}
n^{\left( 1\right)}
&=&\int f^{\left( 0\right)}\mathit{\Phi } ^{\left( 1\right)}\,d {\mathbf{c}}
\nonumber \\*
&=&
n^{\left( 0\right)}\alpha ^{\left( 1,1\right) }
+mn^{\left( 0\right)}{\mathbf{u}}^{\left( 0\right)}\cdot 
\boldsymbol{\alpha }^{\left( 2,1\right)}
\nonumber \\*
&&+\,\frac{1}{2}\,n^{\left( 0\right)}
{\left[ 3kT^{\left( 0\right)}+m{\left( u^{\left( 0\right) }\right)}^2\right]}
\alpha ^{\left( 3,1\right) } ,
\label{r1_condition1}
\\
m{\left( n {\mathbf{u}}\right)}^{\left( {1} \right)}
&=&\int m{\mathbf{c}}f^{\left( 0\right)}\mathit{\Phi } ^{\left( 1\right)}d {\mathbf{c}}
\nonumber \\*
&=&
mn^{\left( 0\right)}{\mathbf{u}}^{\left( {0} \right)}
{\left( \alpha ^{\left( 1,1\right) } + m{\mathbf{u}}^{\left( 0\right)}\cdot 
\boldsymbol{\alpha }^{\left( 2,1\right)}\right)} 
+mn^{\left( 0\right)}{kT^{\left( 0\right)}}
\boldsymbol{\alpha }^{\left( 2,1\right)}
\nonumber \\*
&&+\,\frac{1}{2}\,
mn^{\left( 0\right)}{\mathbf{u}}^{\left( {0} \right)}
{\left[ 5kT^{\left( 0\right)}+m{\left( u^{\left( 0\right) }\right)}^2\right]}
\alpha ^{\left( 3,1\right) } ,
\label{r1_condition2}
\\
\frac{3}{2}\, k\, {\left( n T \right)}^{\left( {1} \right)}
+ \,\frac{1}{2}\, m {\left( n u^{2} \right)}^{\left( {1} \right)}
&=&
\int \frac{1}{2} \, m  c^{2} 
f^{\left( 0\right)}\mathit{\Phi } ^{\left( 1\right)}d {\mathbf{c}}
\nonumber \\*
&=&
\frac{1}{2}\,n^{\left( 0\right)}
{\left[ 3kT^{\left( 0\right)}+m{\left( u^{\left( 0\right) }\right)}^2\right]}
\alpha ^{\left( 1,1\right) }
\nonumber \\*
&&+\,\frac{1}{2}\,m n^{\left( 0\right)}
{\left[ 5kT^{\left( 0\right)}+
m{\left( u^{\left( 0\right)}\right)}^2\right]}
{\mathbf{u}}^{\left( 0\right)}\cdot 
\boldsymbol{\alpha }^{\left( 2,1\right)}
\nonumber \\*
&&+\,\frac{5}{4}\,n^{\left( 0\right)}kT^{\left( 0\right)}
{\left[ 
3kT^{\left( 0\right)}
+2m {\left( u^{\left( 0\right) }\right)}^2\right]}
\alpha ^{\left( 3,1\right) }
\nonumber \\*
&&+\,\frac{1}{4}\,n^{\left( 0\right)}
m^2{u^{\left( 0\right) }}^4
\alpha ^{\left( 3,1\right) }.
\label{r1_condition3}
\end{eqnarray}
In (\ref{r1_condition1})-(\ref{r1_condition3}) vanishing integrals 
(cf. with \cite{chap60}, Chapter~7, \S ~3, Section~1) are neglected.
From (\ref{r1_condition1})-(\ref{r1_condition3}) we have:
\begin{eqnarray}
\alpha ^{\left( 1,1\right) }
&=&\frac{n^{\left( 1\right)}}{n^{\left( 0\right)}}
-\frac{3}{2} \frac{T^{\left( 1\right)}}{T^{\left( 0\right)}} ,
\label{alpha1}
\\
\boldsymbol{\alpha }^{\left( 2,1\right)}
&=&\frac{{\mathbf{u}}^{\left( {1} \right)}}{kT^{\left( 0\right)}} ,
\label{alpha2}
\\
\alpha ^{\left( 3,1\right) }
&=&\frac{1}{kT^{\left( 0\right)}} \frac{T^{\left( 1\right)}}{T^{\left( 0\right)}} .
\label{alpha3}
\end{eqnarray}

To first infinitesimal order terms
(see. \cite{bourbaki2004}, Chapter~V, \S~2, definition 2) expression
\begin{eqnarray}
f^{\left( 0\right)}
+f^{\left( 0\right)} 
\left(
\frac{n^{\left( 1\right)}}{n^{\left( 0\right)}}
-\frac{3}{2} \frac{T^{\left( 1\right)}}{T^{\left( 0\right)}}
+\frac{{\mathbf{u}}^{\left( {1} \right)}}{kT^{\left( 0\right)}}\cdot m\mathbf{C}
+\frac{1}{kT^{\left( 0\right)}} 
\frac{T^{\left( 1\right)}}{T^{\left( 0\right)}}\frac{1}{2}\,mC^{2}
\right) 
\label{f0tilda_expression}
\end{eqnarray}
coincides with the asymptotic expansion of the solution $\tilde{f}^{\left( 0 \right)}$
in the Enskog-Chapman theory
\begin{eqnarray}
\tilde{f}^{\left( 0 \right)}
=n^{\left[ 1\right]}
\left( \frac{m}{2\pi kT^{\left[ 1\right]}}\right) ^{\frac{3}{2}}
e^
{-{\left. m\left( {\mathbf{c}}-
{\mathbf{u}}^{\left[ 1\right]}\right) ^{2} \right/ 
{2kT^{\left[ 1\right]}}}} ,
\label{f0_tilda}
\end{eqnarray}
where 
$n^{\left[ 1\right]}=n^{\left( 0\right)}+n^{\left( 1\right)}$,
$\mathbf{u}^{\left[ 1\right]}=\mathbf{u}^{\left( 0\right)}+
\mathbf{u}^{\left( 1\right)}$ and
$T^{\left[ 1\right]}=T^{\left( 0\right)}+T^{\left( 1\right)}$, 
cf. with Taylor expansion of function $\tilde{f}^{\left( 0 \right)}$
about the point 
$\left( n^{\left( 0\right)}, \mathbf{u}^{\left( 0\right)}, T^{\left( 0\right)}\right)$.
This assertion can be written in the form of:
\begin{eqnarray}
\tilde{f}^{\left( 0 \right)}
\overset{1}{\sim}f^{\left( 0\right)}
+f^{\left( 0\right)} 
\left(
\frac{n^{\left( 1\right)}}{n^{\left( 0\right)}}
-\frac{3}{2} \frac{T^{\left( 1\right)}}{T^{\left( 0\right)}}
+\frac{{\mathbf{u}}^{\left( {1} \right)}}{kT^{\left( 0\right)}}\cdot m\mathbf{C}
+\frac{1}{kT^{\left( 0\right)}} 
\frac{T^{\left( 1\right)}}{T^{\left( 0\right)}}\frac{1}{2}\,mC^{2}
\right)\! .
\label{f0tilda_equivalence}
\end{eqnarray}

Equations (\ref{A_eq}) and (\ref{B_eq}) differ from analogous equations
\cite{chap60}, Chapter~7, (3.9) and (3.10) in the Enskog-Chapman theory
only in use 
$n^{\left( 0\right)}$, $\mathbf{u}^{\left( 0\right)}$ and $T^{\left( 0\right)}$ 
instead of $n$, $\mathbf{u}$ and $T$ (i.e.
$n^{\left[ 1\right]}$, $\mathbf{u}^{\left[ 1\right]}$ and $T^{\left[ 1\right]}$).
In first and second terms in the right-hand side of (\ref{f1_solution}), 
in (\ref{A_expr}) and (\ref{B_expr}) functions 
$n^{\left( 0\right)}$,
$\mathbf{u}^{\left( 0\right)}$ and
$T^{\left( 0\right)}$, as upper in (\ref{r_condition3n}), 
can be, respectively, replaced by functions 
$n^{\left[ 1\right]}$,
$\mathbf{u}^{\left[ 1\right]}$ and
$T^{\left[ 1\right]}$.
Therefore to first infinitesimal order terms expression
\begin{eqnarray}
{f}^{\left( 0 \right)}
\left[
-\frac{1}{n^{\left(0\right) }}\left( \frac{2kT^{\left(0\right) }}{m}\right) 
\mathbf{A}\cdot \frac{\partial \ln T^{\left(0\right) }}{\partial \mathbf{r}}
-\frac{1}{n^{\left(0\right) }}\,\mathrm{B}:
\frac{\partial }{\partial \mathbf{r}}\mathbf{u}^{\left(0\right)}
\right] 
\label{f1n_expression}
\end{eqnarray}
coincides with the solution $\tilde{f}^{\left( 1 \right)}$
in the Enskog-Chapman theory
\begin{eqnarray}
\tilde{f}^{\left( 1 \right)}
=\tilde{f}^{\left( 0 \right)}
\left[
-\frac{1}{n^{\left[ 1\right] }}\left( \frac{2kT^{\left[ 1\right] }}{m}\right) 
\tilde{\mathbf{A}}\cdot \frac{\partial \ln T^{\left[ 1\right] }}{\partial \mathbf{r}}
-\frac{1}{n^{\left[ 1\right] }}\,\tilde{\mathrm{B}}:
\frac{\partial }{\partial \mathbf{r}}\mathbf{u}^{\left[ 1\right]}
\right] . 
\label{f1_tilda}
\end{eqnarray}
Consequently, to first infinitesimal order terms the solution  
${f}^{\left[ 1 \right]}={f}^{\left( 0 \right)}+{f}^{\left( 1 \right)}$
\textit{coincides} with the solution
$\tilde{f}^{\left[ 1 \right]}=\tilde{f}^{\left( 0 \right)}+\tilde{f}^{\left( 1 \right)}$, 
received in the Enskog-Chapman theory:
\begin{eqnarray}
\tilde{f}^{\left[ 1 \right]}
\overset{1}{\sim}{f}^{\left[ 1 \right]} . 
\label{f1tilda_equivalence}
\end{eqnarray}
As a result, with the same exactness 
expressions for heat flux density vector
(\cite{chap60}, Chapter~7, \S ~4)
\begin{eqnarray}
\tilde{\mathbf{q}}^{\left[ 1\right] }
&=&
\tilde{\mathbf{q}}^{\left( 1\right) }
\overset{1}{\sim}\mathbf{q}^{\left[ 1\right] }=\mathbf{q}^{\left( 1\right) }
=\int \frac{1}{2}\,m
\left( {\mathbf{c}}-{\mathbf{u}}^{\left[ 1\right] }\right)^{2}
\left( {\mathbf{c}}-{\mathbf{u}}^{\left[ 1\right] }\right)
f^{\left[ 1\right] }d\mathbf{c}
\nonumber\\*
&\overset{1}{\sim}&
-\frac{2k^{2}T^{\left( 0\right) }}{3m}
\frac{\partial T^{\left( 0\right) }}{\partial \mathbf{r}}
\int \mathbf{A}\cdot I\left( \mathbf{A}\right) d\mathbf{c}
\nonumber\\*
&\overset{1}{\sim}&
-\frac{2k^{2}T^{\left[ 1\right] }}{3m}
\frac{\partial T^{\left[ 1\right] }}{\partial \mathbf{r}}
\int \tilde{\mathbf{A}}\cdot I\left( \tilde{\mathbf{A}}\right) d\mathbf{c}
\nonumber \\*
&=&
-\lambda \frac{\partial T^{\left[ 1\right] }}{\partial \mathbf{r}}
\label{q1}
\end{eqnarray}
and pressure tensor
\begin{eqnarray}
\tilde{{\mathrm{p}}}^{\left[ 1\right] }
&\overset{1}{\sim}&
{\mathrm{p}}^{\left[ 1\right] }
=\int m
\left( {\mathbf{c}}-{\mathbf{u}}^{\left[ 1\right] }\right)
\left( {\mathbf{c}}-{\mathbf{u}}^{\left[ 1\right] }\right)
f^{\left[ 1\right] }d\mathbf{c}
\nonumber\\*
&\overset{1}{\sim}&
\left( 
 n^{\left( 0\right) }kT^{\left( 0\right) }
+n^{\left( 1\right) }kT^{\left( 0\right) }
+n^{\left( 0\right) }kT^{\left( 1\right) }
\right) {\mathrm{U}}
\nonumber\\*
&&
-\frac{1}{5}\,kT^{\left( 0\right) }
\overset{\circ }
{\overline{\overline{\frac{\partial }{\partial \mathbf{r}}
\,\mathbf{u}^{\left( 0\right)}}}}
\int {\mathrm{B}}:I\left( {\mathrm{B}}\right) d\mathbf{c}
\nonumber\\*
&\overset{1}{\sim}&
n^{\left[ 1\right] }kT^{\left[ 1\right] }{\mathrm{U}}
-\frac{1}{5}\,kT^{\left[ 1\right] }
\overset{\circ }
{\overline{\overline{\frac{\partial }{\partial \mathbf{r}}\,\mathbf{u}^{\left[ 1\right] }}}}
\int \tilde{\mathrm{B}}:I\left( \tilde{\mathrm{B}}\right) d\mathbf{c}
\nonumber\\*
&=&
n^{\left[ 1\right] }kT^{\left[ 1\right] }{\mathrm{U}}
-2\mu 
\overset{\circ }
{\overline{\overline{\frac{\partial }{\partial \mathbf{r}}
\,\mathbf{u}^{\left[ 1\right] }}}} ,
\label{p1}
\end{eqnarray}
coincide, 
cf. with the contrary assertion, for example, in
\cite{strum64}, \cite{cercignani75}, \cite{resibois77}. 
In (\ref{p1}) notation is used: 
for arbitrary second-rank tensor ${\mathrm{w}}$
\begin{eqnarray}
{\left( \overline{\overline{\mathrm{w}}}\right)}_{\alpha \beta } 
=\frac{1}{2}\left( w_{\alpha \beta }+w_{\beta \alpha } \right) 
\label{woverline}
\end{eqnarray}
-- corresponding symmetric tensor.

Having solutions $f^{\left( 0\right)}$ and $f^{\left( 1\right)}$,
we can choose: 
one can solve separately gas-dynamic equations systems of 
the first and the second orders
(\ref{trans_m1})-(\ref{trans_e1}) and (\ref{r_eq}), $r=1$, and separately 
find 
$n^{\left( 0\right)}$, $\mathbf{u}^{\left( 0\right)}$, $T^{\left( 0\right)}$ and
$n^{\left( 1\right)}$, $\mathbf{u}^{\left( 1\right)}$, $T^{\left( 1\right)}$,
or, substituting function 
$f^{\left[ 1\right]}=f^{\left( 0\right)}+f^{\left( 1\right)}$ 
in the system of transfer equations 
(or summing systems of equations (\ref{trans_m1})-(\ref{trans_e1}) and 
(\ref{r_eq}), $r=1$), immediately seek solutions 
$n^{\left[ 1\right]}$,
$\mathbf{u}^{\left[ 1\right]}$, 
$T^{\left[ 1\right]}$
of the system (\ref{trans_eq}) of, generally speaking, 
\textit{singularly perturbed} differential equations.

Analogous results can be received for multi-component gas within the
Enskog approach.
These questions together with third order gas-dynamic equations 
system within the Enskog approach will be considered in a next article.

\section{Turbulence as multi-component gas dynamics}
\label{sec:turbulence}

As is well known, laminar flow becomes the turbulent flow, when 
some parameter characterizing the flow, namely, Reynolds number 
\begin{eqnarray}
R=\frac{\rho u L}\mu > 1 .
\label{Reynolds}
\end{eqnarray}
In (\ref{Reynolds}), $\rho $ is the density of gas, $u$ and $L$ are some 
characteristic macroscopic velocity and linear size of the flow, 
$\mu $ is the coefficient of viscosity. 
Having rewritten (\ref{Reynolds}) as 
\begin{eqnarray}
R=\frac{\rho u^{2}}{\mu \frac{u}{L}} ,
\label{Reynolds_frac}
\end{eqnarray}
-- cf. with the expression for viscosity tensor in (\ref{p1}), 
the Reynolds number can be treated as the ratio of 
the macroscopic momentum flux, proportional to $f^{\left( 0 \right)}$, 
to the viscosity-induced microscopic momentum 
flux, proportional to $\tilde{f}^{\left( 1 \right)}$. 
Roughly speaking, viscosity "aligning" the gas molecules according to a 
Maxwellian distribution at the same mean 
velocity and temperature can "process" the microscopic momentum flux 
alone. 
However, if the macroscopic flux outperforms the microscopic, the gas 
flow, necessarily, comes to be stratified to components. 
The flow stratification to components can be also caused by external factors. 

If
\begin{eqnarray}
R \sim \frac{{f^{\left( 0 \right)} }}{{\tilde{f}^{\left( 1 \right)} }} \to \infty , 
\label{Reynolds_paradox}
\end{eqnarray}
then that turbulent flow must be described by the gas-dynamic equations system, 
corresponding to the first approximation order in the approximate method for 
solution of kinetic Boltzmann equation, i.e. without viscosity and heat 
conductivity. 
But the gas-dynamic equations system of the first approximation order 
in the Enskog-Chapman theory can not describe turbulent flow with 
the \textit{entropy increase}, see (\ref{adiabatic}). 
Using the gas-dynamic equations system
(\ref{trans_m})-(\ref{trans_e}) resolves this paradox.

If gas-dynamic equations do not describe  
turbulent gas flows, 
then either something has been missed during the transition from 
the exact solution of the kinetic Boltzmann equation to its approximate 
solution (by the Enskog method) and then to the gas-dynamic 
equations, 
or the kinetic Boltzmann equation does not describe turbulent gas
flows and requires replacement. 
However the last, i.e. necessity of replacement of the kinetic Boltzmann 
equation on another kinetic equation at transition from gas laminar
flow to turbulent gas flow, seems ill-founded.

The gas dynamics of the components with the velocity distribution functions,
close to the Maxwell functions of different mean 
velocities and temperatures, should be described by equations 
(\ref{trans_m})-(\ref{trans_e}). 
From this point of view, 
the observed \textit{stochasticity} of the turbulent flow is similar 
to the stochasticity of the Brownian motion. 
They differ in scale: in the Brownian motion that particle moves 
stochastically, whose mass is comparable to the mass of separate gas 
molecules, while in the turbulent flow that body moves stochastically, 
whose mass is comparable to the mass of separate gas components. 
In (\ref{trans_p})-(\ref{trans_e}) the integral terms 
(proportional to $n_i$, $n_j$) can be  
huge, it explains \textit{unexpected power} of turbulent effects.



\end{document}